\documentclass[final, a4paper, 10pt,twocolumn]{IEEEtran}
\usepackage{epsfig}
\usepackage{amsmath,amssymb}
\usepackage{xcolor}
\usepackage{comment}
\usepackage{algorithm}
\usepackage{algorithmic}
\usepackage{multirow}
\usepackage{slashbox}
\usepackage{rotating}
\usepackage{graphicx}
\usepackage{subfig}
\usepackage{cite}
\usepackage{changebar}
\usepackage{longtable}
\usepackage{chngpage}
\usepackage{array}
\usepackage{booktabs}
\usepackage[printonlyused]{acronym}

\usepackage[style=list,nonumberlist]{glossaries}
\renewcommand{\glsgroupskip}{}
\newglossaryentry{2G}{name=2G, description={second generation}}
\newglossaryentry{3G}{name=3G, description={third generation}}
\newglossaryentry{3GPP}{name=3GPP, description={Third Generation Partnership Project}}
\newglossaryentry{4G}{name=4G, description={fourth generation}}
\newglossaryentry{5G}{name=5G, description={fifth generation}}
\newglossaryentry{2D}{name=2D, description={two-dimensional}}
\newglossaryentry{3D}{name=3D, description={three-dimensional}}

\newglossaryentry{ABC}{name=ABC, description={artificial bee colony}}
\newglossaryentry{ACO}{name=ACO, description={ant colony optimization}}
\newglossaryentry{AF}{name=AF, description={amplify-and-forward}}
\newglossaryentry{AHP}{name=AHP, description={analytical hierarchy process}}
\newglossaryentry{AI}{name=AI, description={artificial intelligence}}
\newglossaryentry{AIS}{name=AIS, description={artificial immune system}}
\newglossaryentry{ANN}{name=ANN, description={artificial neural network}}
\newglossaryentry{AP}{name=AP, description={access point}}
\newglossaryentry{APF}{name=APF, description={artificial potential field}}
\newglossaryentry{APP}{name=APP, description={\textit{a posteriori} probability}}
\newglossaryentry{AWGN}{name=AWGN, description={additive white Gaussian noise}}
\newglossaryentry{AME}{name=AME, description={asymptotic-multiuser-efficiency}}
\newglossaryentry{ASIC}{name=ASIC, description={application-specific integrated circuit}}
\newglossaryentry{A-CPDA}{name=A-CPDA, description={approximate complex-valued probabilistic data association}}
\newglossaryentry{AB-Log-PDA}{name=AB-Log-PDA, description={approximate Bayes' theorem based logarithmic-domain probabilistic data association}}

\newglossaryentry{BER}{name=BER, description={bit-error rate}}
\newglossaryentry{BC-SDPR}{name=BC-SDPR, description={bound-constrained semidefinite programming relaxation}}
\newglossaryentry{BCJR}{name=BCJR, description={Bahl-Cocke-Jelinek-Raviv}}
\newglossaryentry{BICM}{name=BICM, description={bit-interleaved coded modulation}}
\newglossaryentry{BIP}{name=BIP, description={binary integer programming}}
\newglossaryentry{BLER}{name=BLER, description={block-error rate}}
\newglossaryentry{B-PDA}{name=B-PDA, description={bit-based probabilistic data association }}
\newglossaryentry{BPSK}{name=BPSK, description={binary phase-shift keying}}
\newglossaryentry{BS}{name=BS, description={base station}}
\newglossaryentry{BSC}{name=BSC, description={base station controller}}
\newglossaryentry{BALM}{name=BALM, description={block alternating likelihood maximization}}
\newglossaryentry{BQP}{name=BQP, description={Boolean quadratic programming}}
\newglossaryentry{BP}{name=BP, description={belief propagation}}
\newglossaryentry{BI-GDFE}{name=BI-GDFE, description={block-iterative generalized decision feedback equalizer}}
\newglossaryentry{BOA}{name=BOA, description={Bayesian optimization algorithm}}
\newglossaryentry{BCMN/A}{name=BCMN/A, description={broadcasting combined with multi-NACK/ACK}}

\newglossaryentry{CASER}{name=CASER, description={cost-aware secure routing}}
\newglossaryentry{CCI}{name=CCI, description={co-channel interference}}
\newglossaryentry{CCMC}{name=CCMC, description={continuous-input continuous-output memoryless channel}}
\newglossaryentry{CDF}{name=CDF, description={cumulative density function}}
\newglossaryentry{CDM}{name=CDM, description={code-division multiplexing}}
\newglossaryentry{CDMA}{name=CDMA, description={code-division multiple-access}}
\newglossaryentry{CPDA}{name=CPDA, description={complex-valued probabilistic data association}}
\newglossaryentry{CIVA}{name=CIVA, description={centralized immune-Voronoi deployment algorithm}}
\newglossaryentry{CAGR}{name=CAGR, description={compound annual growth rate}}
\newglossaryentry{CMOS}{name=CMOS, description={complementary metal-oxide semiconductor}}

\newglossaryentry{CIR}{name=CIR, description={channel impulse response}}
\newglossaryentry{CP}{name=CP, description={cyclic prefixing}}
\newglossaryentry{CQI}{name=CQI, description={channel quality information}}
\newglossaryentry{CR}{name=CR, description={cognitive radio}}
\newglossaryentry{CR-WSN}{name=CR-WSN, description={cognitive radio aided WSN}}
\newglossaryentry{CSI}{name=CSI, description={channel state information}}
\newglossaryentry{CSIR}{name=CSIR, description={channel state information at the receiver}}
\newglossaryentry{CSIT}{name=CSIT, description={channel state information at the transmitter}}
\newglossaryentry{CLPS}{name=CLPS, description={closest lattice-point search}}
\newglossaryentry{CSPU}{name=CSPU, description={central signal processing unit}}

\newglossaryentry{DAS}{name=DAS, description={distributed antenna system}}
\newglossaryentry{DE}{name=DE, description={differential evolution}}
\newglossaryentry{DF}{name=DF, description={decode-and-forward}}
\newglossaryentry{DFE}{name=DFE, description={decision-feedback equalisation}}
\newglossaryentry{DFD}{name=DFD, description={decision-feedback detector}}
\newglossaryentry{DFT}{name=DFT, description={discrete Fourier transform}}
\newglossaryentry{DID}{name=DID, description={distributed iterative detection}}
\newglossaryentry{DPDA}{name=DPDA, description={distributed probabilistic data association}}
\newglossaryentry{DS-CDMA}{name=DS-CDMA, description={direct-sequence code-division multiple-access}}
\newglossaryentry{DSA}{name=DSA, description={dynamic spectrum access}}
\newglossaryentry{DSC}{name=DSC, description={disjoint set cover}}
\newglossaryentry{DT}{name=DT, description={direct transmission}}
\newglossaryentry{DSNR}{name=DSNR, description={decreasing signal-to-noise ratio}}
\newglossaryentry{DVA-SDPR}{name=DVA-SDPR, description={direct-bit-based virtually antipodal semidefinite programming relaxation}}
\newglossaryentry{DR}{name=DR, description={detectable range}}
\newglossaryentry{DSL}{name=DSL, description={digital subscriber line}}
\newglossaryentry{DQPSK}{name=DQPSK, description={differential quadrature phase-shift keying}}
\newglossaryentry{DMOEA}{name=DMOEA, description={distributed multi-objective evolutionary algorithm}}
\newglossaryentry{DoS}{name=DoS, description={denial-of-service}}
\newglossaryentry{DPAP}{name=DPAP, description={deployment and power assignment problem}}
\newglossaryentry{DEAP}{name=DEAP, description={distributed evolutionary algorithms in Python}}

\newglossaryentry{EGC}{name=EGC, description={equal-gain combining}}
\newglossaryentry{EXIT}{name=EXIT, description={extrinsic information transfer}}
\newglossaryentry{EB-Log-PDA}{name=EB-Log-PDA, description={exact Bayes' theorem based logarithmic-domain probabilistic data association}}
\newglossaryentry{EB}{name=EB, description={exabytes}}
\newglossaryentry{EMA}{name=EMA, description={energy-efficient minimum-latency data aggregation algorithm}}
\newglossaryentry{EHF}{name=EHF, description={extremely high frequency}}
\newglossaryentry{EM}{name=EM, description={expectation-maximization}}
\newglossaryentry{EA}{name=EA, description={evolutionary algorithm}}
\newglossaryentry{EDLA}{name=EDLA, description={energy-density-latency-accuracy}}
\newglossaryentry{EMOCA}{name=EMOCA, description={evolutionary multi-objective crowding algorithm}}
\newglossaryentry{ENS_OR}{name=ENS_OR, description={energy saving via opportunistic routing}}

\newglossaryentry{FA}{name=FA, description={firefly algorithm}}
\newglossaryentry{FLOP}{name=FLOP, description={floating point operation}}
\newglossaryentry{FD}{name=FD, description={frequency-domain}}
\newglossaryentry{FDE}{name=FDE, description={frequency-domain equalisation}}
\newglossaryentry{FD-LE}{name=FD-LE, description={frequency-domain linear equalisation}}
\newglossaryentry{FD-DFE}{name=FD-DFE, description={frequency-domain decision-feedback equalisation}}
\newglossaryentry{FDM}{name=FDM, description={frequency-division multiplexing}}
\newglossaryentry{FDMA}{name=FDMA, description={frequency-division multiple-access}}
\newglossaryentry{FEC}{name=FEC, description={forward-error-correction}}
\newglossaryentry{FR}{name=FR, description={frequency reuse}}
\newglossaryentry{FH-CDMA}{name=FH-CDMA, description={frequency-hopped code-division multiple-access}}
\newglossaryentry{FIR}{name=FIR, description={finite impulse response}}
\newglossaryentry{FCSD}{name=FCSD, description={fixed-complexity sphere decoding/decoder}}
\newglossaryentry{FER}{name=FER, description={frame-error rate}}
\newglossaryentry{FL}{name=FL, description={fuzzy logic}}
\newglossaryentry{FRMOO}{name=FRMOO, description={fuzzy random multi-objective optimization}}

\newglossaryentry{GSNR}{name=GSNR, description={greatest signal-to-noise ratio}}
\newglossaryentry{GA}{name=GA, description={genetic algorithm}}
\newglossaryentry{GMOP}{name=GMOP, description={general multi-objective program}}
\newglossaryentry{GP}{name=GP, description={goal programming}}
\newglossaryentry{GAF}{name=GAF, description={geographical adaptive fidelity}}
\newglossaryentry{GUIMOO}{name=GUIMOO, description={graphical user interface for multi-objective optimization}}

\newglossaryentry{HSPA}{name=HSPA, description={high speed packet access}}
\newglossaryentry{HNN}{name=HNN, description={Hopfield neural network}}
\newglossaryentry{HBOA}{name=HBOA, description={hierarchical Bayesian optimization algorithm}}

\newglossaryentry{IDSS}{name=IDSS, description={intelligent decision support system}}
\newglossaryentry{IC}{name=IC, description={integrated circuit}}
\newglossaryentry{ICA}{name=ICA, description={imperialist competitive algorithm}}
\newglossaryentry{ICI}{name=ICI, description={interchannel interference}}
\newglossaryentry{ICT}{name=ICT, description={information and communication technology}}
\newglossaryentry{ID}{name=ID, description={iterative decoding}}
\newglossaryentry{IDD}{name=IDD, description={iterative detection and decoding}}
\newglossaryentry{IDDR}{name=IDDR, description={integrity and delay differentiated routing}}
\newglossaryentry{IDFT}{name=IDFT, description={inverse discrete Fourier transform}}
\newglossaryentry{ISI}{name=ISI, description={intersymbol interference}}
\newglossaryentry{ISM}{name=ISM, description={industrial, scientific, and medical}}
\newglossaryentry{IMSE}{name=IMSE, description={increasing mean-square error}}
\newglossaryentry{IEEE}{name=IEEE, description={Institute of Electrical and Electronics Engineers}}
\newglossaryentry{IVA-SDPR}{name=IVA-SDPR, description={index-bit-based virtually antipodal semidefinite programming relaxation}}
\newglossaryentry{IAI}{name=IAI, description={interantenna interference}}
\newglossaryentry{IPA}{name=IPA, description={interior point algorithm}}
\newglossaryentry{IVD}{name=IVD, description={\textit{in vivo} diagnostic}}
\newglossaryentry{IoT}{name=IoT, description={Internet of Things}}

\newglossaryentry{JPDA}{name=JPDA, description={joint probabilistic data association}}
\newglossaryentry{JMLD}{name=JMLD, description={joint maximum likelihood detection}}
\newglossaryentry{JD}{name=JD, description={joint detection}}

\newglossaryentry{LAS}{name=LAS, description={likelihood ascent search}}
\newglossaryentry{LE}{name=LE, description={linear equalisation}}
\newglossaryentry{LLR}{name=LLR, description={logarithmic likelihood ratio}}
\newglossaryentry{Log-MAP}{name=Log-MAP, description={logarithmic maximum a-posteriori probability}}
\newglossaryentry{LP}{name=LP, description={linear programming}}
\newglossaryentry{LTE}{name=LTE, description={long-term evolution}}
\newglossaryentry{LTE-A}{name=LTE-A, description={Long Term Evolution-Advanced}}
\newglossaryentry{LTI}{name=LTI, description={linear time-invariant}}
\newglossaryentry{LS}{name=LS, description={least-squares}}
\newglossaryentry{LS-MIMO}{name=LS-MIMO, description={large-scale multiple-input multiple-output}}
\newglossaryentry{LMSE}{name=LMSE, description={least mean-square error}}
\newglossaryentry{LDPC}{name=LDPC, description={low-density parity-check}}
\newglossaryentry{LMR}{name=LMR, description={linear matrix representation}}
\newglossaryentry{LSD}{name=LSD, description={list sphere decoding}}
\newglossaryentry{LSPI}{name=LSPI, description={least squares policy iteration}}
\newglossaryentry{LMI}{name=LMI, description={linear matrix inequality}}
\newglossaryentry{LZF}{name=LZF, description={linear zero-forcing}}
\newglossaryentry{LR}{name=LR, description={lattice-reduction}}
\newglossaryentry{LLL}{name=LLL, description={Lenstra-Lenstra-Lov\'{a}sz}}

\newglossaryentry{MA}{name=MA, description={memetic algorithm}}
\newglossaryentry{MAC}{name=MAC, description={medium access control}}
\newglossaryentry{MAP}{name=MAP, description={maximum \textit{a posteriori}}}
\newglossaryentry{MBER}{name=MBER, description={minimum bit error rate}}
\newglossaryentry{MC-CDMA}{name=MC-CDMA, description={multicarrier code-division multiple-access}}
\newglossaryentry{MCP}{name=MCP, description={multi-cell processing}}
\newglossaryentry{MDP}{name=MDP, description={Markov decision process}}
\newglossaryentry{MF}{name=MF, description={matched filter}}
\newglossaryentry{MFB}{name=MFB, description={matched filter bound}}
\newglossaryentry{MIMO}{name=MIMO, description={multiple-input multiple-output}}
\newglossaryentry{MISO}{name=MISO, description={multiple-input single-output}}
\newglossaryentry{ML}{name=ML, description={maximum likelihood}}
\newglossaryentry{MLSE}{name=MLSE, description={maximum likelihood sequence estimator/estimation}}
\newglossaryentry{MIC}{name=MIC, description={multistage interference cancellation}}
\newglossaryentry{MMSE}{name=MMSE, description={minimum mean-square error}}
\newglossaryentry{MMF}{name=MMF, description={multimode fibre}}
\newglossaryentry{MAME}{name=MAME, description={maximum asymptotic-multiuser-efficiency}}
\newglossaryentry{MRC}{name=MRC, description={maximum ratio combining}}
\newglossaryentry{MSE}{name=MSE, description={mean-square error}}
\newglossaryentry{MS}{name=MS, description={mobile station}}
\newglossaryentry{MUD}{name=MUD, description={multiuser detection/detector}}
\newglossaryentry{MUI}{name=MUI, description={multiuser interference}}
\newglossaryentry{M2M}{name=M2M, description={machine-to-machine}}
\newglossaryentry{MFSK}{name=MFSK, description={multiple frequency-shift keying}}
\newglossaryentry{MAI}{name=MAI, description={multiple-access interference}}
\newglossaryentry{MSI}{name=MSI, description={multiple-stream interference}}
\newglossaryentry{MSDD}{name=MSDD, description={multi-symbol differential detection/detector}}
\newglossaryentry{MMW}{name=MMW, description={millimetre wave}}
\newglossaryentry{MGS}{name=MGS, description={mixed Gibbs sampling}}
\newglossaryentry{MR}{name=MR, description={multiple restart}}
\newglossaryentry{MSDSD}{name=MSDSD, description={multiple symbol differential sphere decoder}}
\newglossaryentry{MED}{name=MED, description={minimum Euclidean distance}}
\newglossaryentry{MOGA}{name=MOGA, description={multi-objective genetic algorithm}}
\newglossaryentry{MGA}{name=MGA, description={micro-genetic algorithm}}
\newglossaryentry{MODA}{name=MODA, description={multi-objective deployment algorithm}}
\newglossaryentry{MODE}{name=MODE, description={multi-objective differential evolution}}
\newglossaryentry{MOEA}{name=MOEA, description={multi-objective evolutionary algorithm}}
\newglossaryentry{MOEA/D}{name=MOEA/D, description={multi-objective evolutionary algorithm based on decomposition}}
\newglossaryentry{MOEA/DFD}{name=MOEA/DFD, description={multi-objective evolutionary algorithm based on decomposition with fuzzy dominance}}
\newglossaryentry{MOGLS}{name=MOGLS, description={multi-objective genetic local search}}
\newglossaryentry{MOICA}{name=MOICA, description={multi-objective imperialist competitive algorithm}}
\newglossaryentry{MOMGA}{name=MOMGA, description={multi-objective messy genetic algorithm}}
\newglossaryentry{MOMGA-II}{name=MOMGA-II, description={multi-objective messy genetic algorithm-II}}
\newglossaryentry{MOO}{name=MOO, description={multi-objective optimization}}
\newglossaryentry{MOP}{name=MOP, description={multi-objective optimization problem}}
\newglossaryentry{MOSS}{name=MOSS, description={multi-objective scatter search}}
\newglossaryentry{MOTS}{name=MOTS, description={multi-objective tabu search}}
\newglossaryentry{MWSNS}{name=MWSNS, description={mobile wireless sensor networks}}
\newglossaryentry{MOMHLib++}{name=MOMHLib++, description={multiple objective metaheuristics library in C++}}

\newglossaryentry{NPGA}{name=NPGA, description={niched Pareto genetic algorithm}}
\newglossaryentry{NSGA}{name=NSGA, description={non-dominated sorting genetic algorithm}}
\newglossaryentry{NSGA-II}{name=NSGA-II, description={non-dominated sorting genetic algorithm-II}}
\newglossaryentry{NUM}{name=NUM, description={network utility maximization}}
\newglossaryentry{NSC}{name=NSC, description={non-systematic convolutional}}
\newglossaryentry{NP}{name=NP-hard, description={nondeterministic polynomial-time}}
\newglossaryentry{NP-hard}{name=NP-hard, description={nondeterministic polynomial-time hard}}
\newglossaryentry{NP-complete}{name=NP-complete, description={nondeterministic polynomial-time complete}}

\newglossaryentry{OMOEA}{name=OMOEA, description={orthogonal multi-objective evolutionary algorithm}}
\newglossaryentry{OFDM}{name=OFDM, description={orthogonal frequency-division multiplexing}}
\newglossaryentry{OFDMA}{name=OFDMA, description={orthogonal frequency-division multiple-access}}
\newglossaryentry{OSIC}{name=OSIC, description={ordered successive interference cancellation}}
\newglossaryentry{OSI}{name=OSI, description={open systems interconnection}}
\newglossaryentry{OVRP-TD}{name=OVRP-TD, description={open vehicle routing problems with time deadlines}}

\newglossaryentry{PDA}{name=PDA, description={probabilistic data association}}
\newglossaryentry{PDF}{name=PDF, description={probability density function}}
\newglossaryentry{PE}{name=PE, description={partial equalisation}}
\newglossaryentry{PHY}{name=PHY, description={physical layer}}
\newglossaryentry{PIC}{name=PIC, description={parallel interference cancellation}}
\newglossaryentry{P/S}{name=P/S, description={parallel-to-serial}}
\newglossaryentry{PAM}{name=PAM, description={pulse-amplitude modulation}}
\newglossaryentry{PI-SDPR}{name=PI-SDPR, description={polynomial-inspired semidefinite programming relaxation}}
\newglossaryentry{PSD}{name=PSD, description={positive semidefinite}}
\newglossaryentry{PSO}{name=PSO, description={particle swarm optimization}}
\newglossaryentry{PD-IPA}{name=PD-IPA, description={primal-dual interior-point algorithm}}
\newglossaryentry{PSK}{name=PSK, description={phase-shift keying}}
\newglossaryentry{PER}{name=PER, description={packet-error rate}}
\newglossaryentry{PBBF}{name=PBBF, description={probability-based broadcast forwarding}}
\newglossaryentry{PF}{name=PF, description={Pareto front}}
\newglossaryentry{PS}{name=PS, description={Pareto set}}
\newglossaryentry{PAES}{name=PAES, description={Pareto archive evolution strategy}}
\newglossaryentry{PESA}{name=PESA, description={Pareto envelope-based selection algorithm}}
\newglossaryentry{PESA-II}{name=PESA-II, description={Pareto envelope-based selection algorithm-II}}
\newglossaryentry{PTW}{name=PTW, description={pipelined tone wake-up}}

\newglossaryentry{QoS}{name=QoS, description={quality-of-service}}
\newglossaryentry{QAM}{name=QAM, description={quadrature amplitude modulation}}
\newglossaryentry{QPSK}{name=QPSK, description={quadrature phase-shift keying}}
\newglossaryentry{QRD}{name=QRD, description={QR-decomposition}}

\newglossaryentry{RF}{name=RF, description={radio frequency}}
\newglossaryentry{RL}{name=RL, description={reinforcement learning}}
\newglossaryentry{RSC}{name=RSC, description={recursive systematic convolutional}}
\newglossaryentry{RPDA}{name=RPDA, description={real-valued probabilistic data association}}
\newglossaryentry{RTS}{name=RTS, description={reactive tabu search}}
\newglossaryentry{R-MCMC}{name=R-MCMC, description={randomized Markov chain Monte Carlo}}
\newglossaryentry{RS}{name=RS, description={randomized search}}

\newglossaryentry{SA}{name=SA, description={simulated annealing}}
\newglossaryentry{SC}{name=SC, description={soft combining}}
\newglossaryentry{SD}{name=SD, description={stochastic diffusion}}
\newglossaryentry{SDD}{name=SDD, description={subgradient dual decomposition}}
\newglossaryentry{SEDR}{name=SEDR, description={security and energy-efficient disjoint routing}}
\newglossaryentry{SIOA}{name=SIOA, description={swarm intelligence based optimization algorithm}}
\newglossaryentry{SPEA}{name=SPEA, description={strength Pareto evolutionary algorithm}}
\newglossaryentry{SPEA2}{name=SPEA2, description={strength Pareto evolutionary algorithm-2}}
\newglossaryentry{S/P}{name=S/P, description={serial-to-parallel}}
\newglossaryentry{SC-FDE}{name=SC-FDE, description={single-carrier frequency-domain equalisation}}
\newglossaryentry{SC-FDMA}{name=SC-FDMA, description={single-carrier frequency-division multiple-access}}
\newglossaryentry{SDM}{name=SDM, description={space-division multiplexing}}
\newglossaryentry{SDMA}{name=SDMA, description={space-division multiple-access}}
\newglossaryentry{SDP}{name=SDP, description={semidefinite programming}}
\newglossaryentry{SDPR}{name=SDPR, description={semidefinite programming relaxation}}
\newglossaryentry{SIC}{name=SIC, description={successive interference cancellation}}
\newglossaryentry{SISO}{name=SISO, description={soft-input soft-output}}
\newglossaryentry{SIMO}{name=SIMO, description={single-input multiple-output}}
\newglossaryentry{SINR}{name=SINR, description={signal-to-interference-plus-noise ratio}}
\newglossaryentry{SIR}{name=SIR, description={signal-to-interference ratio}}
\newglossaryentry{SM}{name=SM, description={spatial multiplexing}}
\newglossaryentry{SNR}{name=SNR, description={signal-to-noise ratio}}
\newglossaryentry{SP}{name=SP, description={set partitioning}}
\newglossaryentry{SUMF}{name=SUMF, description={single-user matched filter}}
\newglossaryentry{SE}{name=SE, description={Schnorr-Euchner}}
\newglossaryentry{STBC}{name=STBC, description={space-time block code/coded}}
\newglossaryentry{SER}{name=SER, description={symbol-error rate}}
\newglossaryentry{SUD}{name=SUD, description={single-user detection}}
\newglossaryentry{SAIC}{name=SAIC, description={single-antenna interference cancellation}}
\newglossaryentry{SUMIS}{name=SUMIS, description={subspace marginalization aided interference suppression}}
\newglossaryentry{STEM}{name=STEM, description={sparse topology and energy management}}

\newglossaryentry{TCM}{name=TCM, description={trellis-coded modulation}}
\newglossaryentry{TPSMA}{name=TPSMA, description={territorial predator scent marking algorithm}}
\newglossaryentry{TTCM}{name=TTCM, description={turbo trellis-coded modulation}}
\newglossaryentry{TDD}{name=TDD, description={time-division-duplex}}
\newglossaryentry{TDM}{name=TDM, description={time-division multiplexing}}
\newglossaryentry{TDMA}{name=TDMA, description={time-division multiple-access}}

\newglossaryentry{UMR}{name=UMR, description={unified matrix representation}}
\newglossaryentry{UEP}{name=UEP, description={unequal error protection}}
\newglossaryentry{UE}{name=UE, description={user equipment}}
\newglossaryentry{UCS}{name=UCS, description={unified-client-server}}

\newglossaryentry{VA}{name=VA, description={virtually antipodal}}
\newglossaryentry{VBLAST}{name=VBLAST, description={vertical Bell Laboratories layered space-time}}
\newglossaryentry{VB}{name=VB, description={Viterbo-Boutros}}
\newglossaryentry{VLSI}{name=VLSI, description={very-large-scale integration}}
\newglossaryentry{VA-SDPR}{name=VA-SDPR, description={virtually antipodal semidefinite programming relaxation}}
\newglossaryentry{VNI}{name=VNI, description={visual network index}}
\newglossaryentry{VER}{name=VER, description={vector-error rate}}

\newglossaryentry{WBAN}{name=WBAN, description={wireless body area network}}
\newglossaryentry{WLS}{name=WLS, description={weighted least-squares}}
\newglossaryentry{WiMAX}{name=WiMAX, description={Worldwide Interoperability for Microwave Access}}
\newglossaryentry{WSN}{name=WSN, description={wireless sensor network}}
\newglossaryentry{WLAN}{name=WLAN, description={wireless local area network}}

\newglossaryentry{XOR}{name=XOR, description={exclusive or}}

\newglossaryentry{ZP}{name=ZP, description={zero-padding}}
\newglossaryentry{ZF}{name=ZF, description={zero-forcing}} 
\makeglossaries

\begin{document}
\title{A Survey of Multi-Objective Optimization in Wireless Sensor Networks: Metrics, Algorithms and Open Problems}

\author{Zesong Fei,~\IEEEmembership{Senior Member,~IEEE,} Bin Li, Shaoshi Yang,~\IEEEmembership{Member,~IEEE,} \\  Chengwen Xing,~\IEEEmembership{Member,~IEEE,} Hongbin Chen, and Lajos~Hanzo,~\IEEEmembership{Fellow,~IEEE}

\thanks{The financial support of the National Natural Science Foundation of China (Grant No. 61371075 and 61421001), of the 111 Project of China (Grant No. B14010), and of the European Research Council (ERC) Advanced Fellow Grant ``Beam-Me-Up'' is gratefully acknowledged.
}

\thanks{Z. Fei, B. Li, and C. Xing are with the School of Information and Electronics, Beijing Institute of Technology, Beijing, 100081, China (e-mail: feizesong@bit.edu.cn; libin$\_$sun@bit.edu.cn; chengwenxing@ieee.org).}
\thanks{S. Yang and L. Hanzo are with the School of Electronics and Computer Science, University of Southampton, Southampton, SO17 1BJ, UK (e-mail: \{sy7g09, lh\}@ecs.soton.ac.uk).}
\thanks{H. Chen is with the Key Laboratory of Cognitive Radio and Information Processing (Ministry of Education), Guilin University of Electronic Technology, Guilin, 541004, China (email: chbscut@guet.edu.cn).}
}

\markboth{Accepted to appear on IEEE Communications Surveys \& Tutorials, Sept. 2016.}%
{Shell \MakeLowercase{\textit{et al.}}: Bare Demo of IEEEtran.cls
for Journals}

\maketitle

%%===============================Abstract==============================%%
\begin{abstract}
%\boldmath
Wireless sensor networks (WSNs) have attracted substantial research interest, especially in the context of performing monitoring and surveillance tasks. However, it is challenging to strike compelling trade-offs amongst the various conflicting optimization criteria, such as the network's energy dissipation, packet-loss rate, coverage and lifetime. This paper provides a tutorial and survey of recent research and development efforts addressing this issue by using the technique of multi-objective optimization (MOO). First, we provide an overview of the main optimization objectives used in WSNs. Then, we elaborate on various prevalent approaches conceived for MOO, such as the family of mathematical programming based scalarization methods, the family of heuristics/metaheuristics based optimization algorithms, and a variety of other advanced optimization techniques. Furthermore, we summarize a range of recent studies of MOO in the context of WSNs, which are intended to provide useful guidelines for researchers to understand the referenced literature. Finally, we discuss a range of open problems to be tackled by future research.
\end{abstract}
%%========================Keywords===============================%%

%%========================acronym===============================%%
\IEEEpeerreviewmaketitle

{\footnotesize{\printglossaries}}

\par

%%========================Section I===============================%%
\section{Introduction \label{sec1}}
\subsection{Motivation}
\IEEEPARstart{W}{ireless} sensor networks (\glspl{WSN}) consist of a large number of compact, low-cost, low-power, multi-functional sensor nodes that communicate wirelessly over short distances \cite{A11,Yick_Computer2008}. In WSNs, the sensor nodes are generally deployed randomly in the field of interest, which are extensively used for performing monitoring and surveillance tasks \cite{Bruckner_TII2012,Yetgin_Access2015,Han_Magazine2015}. Depending on the specific application scenarios, WSNs may rely on diverse performance metrics to be optimized. For example, the energy efficiency and network lifetime are among the major concerns in WSNs, since the sensor nodes are typically powered by battery, whose replacement is often difficult. Furthermore, the network coverage, latency and the fairness among sensor nodes are important for maintaining the quality-of-service (\gls{QoS}) \cite{G5,Cheng_Parallel2014}. In practice, these metrics often conflict with each other, hence the careful balancing of the trade-offs among them is vital in terms of optimizing the overall performance of WSNs in real applications.

In conventional WSN designs, typically the most salient performance metric is chosen as the optimization objective, while the remaining performance metrics are normally treated as the constraints of the optimization problem. Such single-objective optimization approaches, however, may be unfair and unreasonable in real WSN applications, since it artificially over-emphasizes the importance of one of the metrics to the detriment of the rest \cite{X1}.
Hence, a more realistic optimization is to simultaneously satisfy multiple objectives, such as the maximal energy efficiency, the shortest delay, the longest network lifetime, the highest reliability, and the most balanced distribution of the nodes' residual energy, or the trade-offs among the above objectives\cite{Marler_Optimization2004,Tharmarasa_Systems2009}. Accordingly, multi-objective optimization (\gls{MOO}) can be naturally adopted for solving the above problem, since it may be more consistent with the realistic scenarios \cite{M2}.

MOO algorithms have been a subject of intense interest to researchers for solving diverse multi-objective optimization problems (\glspl{MOP}), in which multiple  objectives are treated simultaneously subject to a set of constraints \cite{D01}. However, it is infeasible for multiple objectives to achieve their respective optima at the same time, thus there may not exist a single globally optimal solution, which is the best with respect to all objectives.
Nevertheless, there exists a set of Pareto-optimal or non-dominated solutions generating a set of Pareto-optimal outcomes/objective vectors, which is called Pareto front/frontier (\gls{PF}) or Pareto boundary/curve/surface. Explicitly, the PF is generated by the specific set of solutions, for which none of the multiple objectives can be improved without sacrificing the other objectives \cite{T1}. This set of Pareto-optimal or non-dominated solutions constitutes the focus of our interest, and it is also called the Pareto-efficient set or Pareto set (\gls{PS}) that is mapped to the PF in the objective function space \cite{Y1}.

Diverse approaches, such as mathematical programming based scalarization methods and nature-inspired metaheuristics, may be used for finding the PSs of MOPs. Scalarizing an MOP means formulating a single-objective optimization problem such that optimal solutions to the single-objective optimization problem are Pareto-optimal solutions to the MOP\cite{Ehrgott2005}. In addition, it is often required that every Pareto-optimal solution can be reached with the aid of specific parameters of the scalarization. Representatives of scalarization methods include the linear weighted-sum method, the $\boldsymbol \varepsilon$-constraints method\cite{Ehrgott2005} and goal programming (\gls{GP}) based methods. MOPs are more often solved by bio-inspired metaheuristics, such as multi-objective evolutionary algorithms (\glspl{MOEA}) \cite{T9,T3} and swarm intelligence based optimization algorithms (\glspl{SIOA}) \cite{Y2}. MOEAs aim for finding a set of representative Pareto-optimal solutions in a single run \cite{Y1,Zhou_Swarm2011,Tan_Review2002}.
As a subset of MOEAs, the multi-objective genetic algorithms (\glspl{MOGA}), such as the strength Pareto evolutionary algorithm (\gls{SPEA}) \cite{T9} and the non-dominated sorting genetic algorithm-II (\gls{NSGA-II}) \cite{D13}, have been particularly widely researched in the family of MOO algorithms\cite{Coello_ACM2000}, because they are capable of efficiently constructing an approximate PF. This is mainly due to the fact that MOGAs accommodate a diverse variety of bio-inspired operators to iteratively generate a population of feasible solutions.
Compared to genetic algorithms (\glspl{GA}) that rely on the interplay between genetics and biological evolution, SIOAs seek to understand the collective behavior of animals, particularly insects, and to use this understanding for solving complex, nonlinear problems. One of the most widely used SIOAs is the ant colony optimization (\gls{ACO}) algorithm \cite{K1}, which has indeed been invoked for solving the MOPs in WSNs \cite{T24}.
Several other bio-inspired algorithms related to swarm intelligence will be surveyed in Section \ref{sec5}.

\subsection{Contributions of This Survey}
In this paper, we focus our attention on various basic concepts, conflicting performance criteria/optimization objectives, as well as the MOO techniques conceived for striking a trade-off in the context of WSNs.
The contributions of our work are four-fold, which are listed as follows:
\begin{itemize}
\item We provide in-depth discussions on the basics, metrics and relevant algorithms conceived for MOO in WSNs.
\item We present a comprehensive coverage and clear classification of various prevalent MOO algorithms conceived for solving MOPs, and clarify the strengths and weaknesses of each MOO algorithm in the context of WSNs.
\item We provide an exhaustive review of the up-to-date research progress of MOO in WSNs according to different trade-off metrics.
\item We highlight a variety of open research challenges and identify possible future trends for MOO in WSNs, according to the latest developments of WSNs.
\end{itemize}
\begin{figure*}[tbp]
\centering
\includegraphics[width=16cm]{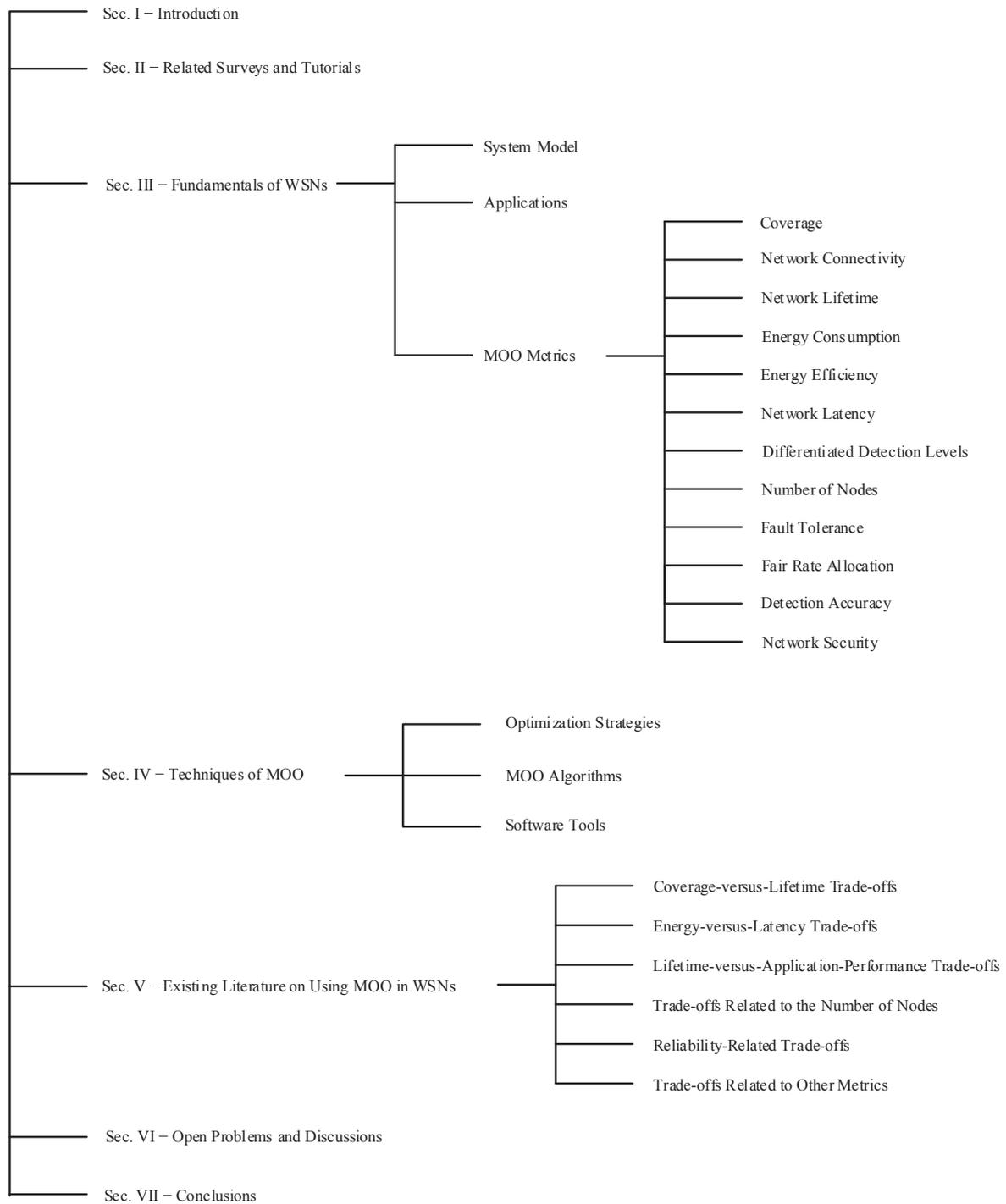}
\caption{The organization of this paper.}
\label{organization}
\end{figure*}

\subsection{Paper Organization}
The reminder of this paper is organized as follows. In Section \ref{sec2}, we summarize the related surveys of MOO in WSNs. In Section \ref{sec3}, we commence with an overview of WSNs in terms of their system model and applications. Furthermore, we introduce the main optimization objectives of interest in WSNs.
% %In Section \ref{sec4}, we introduce the main optimization objectives of interest in WSNs.
In Section \ref{sec5}, we present the family of MOO techniques that can in principle be used for solving this kind of problems. In Section \ref{sec6}, we provide an overview of the existing studies dedicated to multi-objective methods in WSNs. Finally, in Section \ref{sec7} we describe a range of open problems and possible future research directions, followed by our conclusions in Section \ref{sec8}. For the sake of explicit clarity, the organization of this paper is shown in Fig. \ref{organization}.

\section{Related Surveys and Tutorials \label{sec2}}
A range of surveys have been dedicated to diverse single-objective research domains in WSNs, such as their energy efficiency \cite{G1}, routing \cite{G2}, congestion control \cite{G3}, their MAC protocols \cite{G4}, data collection \cite{G5}, privacy and security \cite{G6}, localization
\cite{G7}, \cite{G8}, cross-layer QoS guarantees\cite{Anbagi_Surveys2016}, sink mobility management \cite{Gu_Surveys2016}, and network virtualization \cite{Khan_survey2016}.

In recent years, several surveys and tutorials advocated MOO methods for optimizing the conflicting performance objectives of WSNs. Specifically, the authors of\cite{D12} provided a review of recent studies on multi-objective scheduling and discussed its future research trends. In \cite{R1}, the MOO criteria and strategies conceived for node deployment in WSNs were surveyed. Performance trade-off mechanisms of the routing protocols designed for energy-efficient WSNs were reviewed in \cite{Gao2016}, where various artificial intelligence techniques and the related technical features of the routing protocols were discussed.
The authors of \cite{Coello_Magazine2006} surveyed the most representative MOEAs and their major applications from a historical perspective.
Konak \textit{et al.}\cite{Konak_tutorial2006} presented a comprehensive survey and tutorial of MOGAs. Furthermore, Adnan \textit{et al.}\cite{B11} provided a holistic overview of bio-inspired optimization techniques, such as particle swarm optimization (\gls{PSO}), ACO and GA. In particular, the authors of \cite{Kulkarni_PartC2011} presented a brief survey of how to apply PSO in WSN applications, while bearing in mind the peculiar characteristics of sensor nodes. They also presented a state-of-the-art survey of computational intelligence in the context of WSNs and highlighted numerous challenges facing each of the MOPs discussed\cite{Kulkarni2011_Tutorials}.
Additionally, the authors of \cite{Jabbar2013} reviewed the optimization in biological systems and discussed bio-inspired optimization of non-biological systems. In contrast to other surveys, the authors of \cite{T41} provided a classification of algorithms proposed in the literature for planned deployment of WSNs. They discussed and compared diverse WSN deployment algorithms in terms of their assumptions, objectives and performance. Additionally,
in a more recent study \cite{Iqbal2015,Iqbal_applications2016} the authors reviewed the MOO techniques and simulation tools conceived for solving different problems related to the design, operation, deployment, placement, planning and management of WSNs. The above-mentioned surveys related to MOO in WSNs are outlined at a glance in Table \ref{relatedworks}, which allows the readers to capture the main contributions of each of the existing surveys.

\begin{table*}\small
\centering
\setlength{\tabcolsep}{2pt}
\renewcommand{\arraystretch}{1.3}
\extrarowheight 3pt
\caption{Existing Surveys Relating to MOO in WSNs.}
\begin{tabular}{| l | p{6cm} | l | p{7cm} |}
\hline
\textbf{Reference} & \textbf{Focus Topics} & \textbf{Reference} & \textbf{Focus Topics} \\
\hline
\hline
\cite{D12} & multi-objective scheduling & \cite{Kulkarni_PartC2011}  & a brief survey of PSO \\
\hline
\cite{R1} & sensor node deployment &  \cite{Kulkarni2011_Tutorials}   & computational intelligence paradigms \\
\hline
\cite{Gao2016} & artificial intelligence methods for routing protocols  & \cite{Jabbar2013}   & analogy between optimization in biological systems and bio-inspired optimization in non-biological systems\\
\hline
\cite{Coello_Magazine2006}  & MOEAs & \cite{T41}   & multi-objective node deployment algorithms \\
\hline
\cite{Konak_tutorial2006}  & MOGAs & \cite{Iqbal2015}  & MOO techniques associated with the design, operation, deployment, placement, planning and management  \\
\hline
\cite{B11}  & bio-inspired optimization techniques & \cite{Iqbal_applications2016}   & engineering applications and simulation tools \\
\hline
\end{tabular}
\label{relatedworks}
\end{table*}

%Although existing surveys on multi-objective paradigm of WSNs can be found in \cite{D12} and \cite{Iqbal_applications2016}, these excellent surveys have not properly classified and surveyed on the subject of balancing multi-objective trade-offs.
%Our survey is much more general by including optimization objectives, optimization strategies, and optimization algorithms.
%We classify these optimization algorithms in the open literature for MOO of WSNs more concretely and more subtly.
%Explicitly, we discuss the strengths and weaknesses of each MOO approach making a comparison between them.
%Furthermore, we survey a great deal of up-to-date available literature on MOO in WSNs.

\section{Fundamentals of WSNs \label{sec3}}

\subsection{System Model}
\begin{figure}[tbp]
\centering
\includegraphics[width=2.5in]{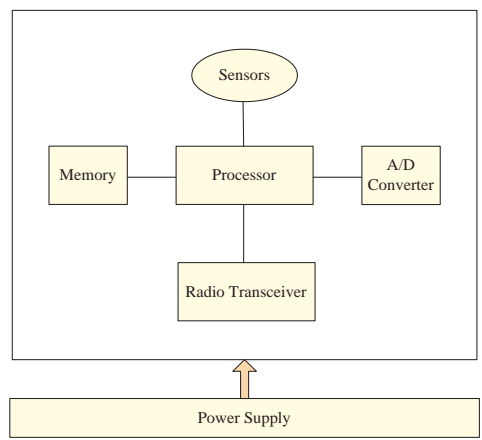}
\caption{Typical architecture of a WSN node.}
\label{sys1}
\end{figure}

WSNs generally consist of hundreds or potentially even thousands of spatially distributed, low-cost, low-power, multi-functional, autonomous sensor nodes and communicate over short distances \cite{A11}. Each node is usually equipped with a sensor unit, a processor, a radio transceiver, an A/D converter, a memory unit, and a power supply (battery). The typical architecture of a WSN node is illustrated in Fig. \ref{sys1}. A WSN node may also have additional application-dependent components attached, such as the location finding system and mobilizer. By combining these different
components into a miniaturized device, these sensor nodes become multi-functional. In other words, the structure and characteristics of sensor nodes depend both on their electronic, mechanical and communication limitations, as well as on their application-specific requirements. One of the great challenges facing WSNs is to use such resource-constrained sensor nodes to meet certain application requirements, including sensing coverage, network lifetime and end-to-end delay.

\begin{figure*}[t]
\centering
\includegraphics[width=4.0in]{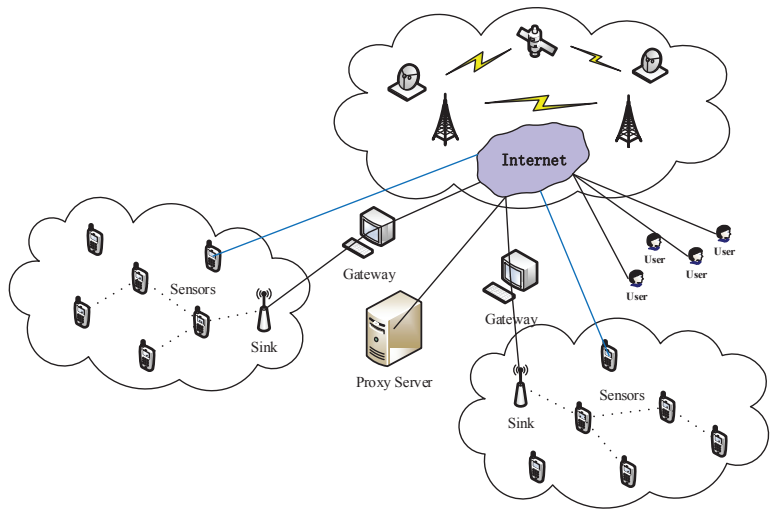}
\caption{Wireless sensor networks and their relationship to infrastructure-based networks.}
\label{sys2}
\end{figure*}

Typically, sensor nodes are grouped into clusters, and each cluster has a node that acts as the cluster head, which has more resources and computational power than the other cluster nodes. All nodes gather and deliver their sensed information to the cluster head, which in turn forwards it to a specialized node, namely the sink node or base station, via a hop-by-hop wireless communication link. In indoor scenarios, a WSN is typically rather small and consists of a single cluster supported by a single base station. Multiple clusters relying on multiple base stations are possible in a large-scale deployment of WSNs.
Fig. \ref{sys2} shows the relationship between WSNs and the infrastructure-based networks. Typically, a sink node or base station is responsible for gathering the uplink information gleaned from sensor nodes through either single-hop or multi-hop communications. Then, the sink node sends the collected information to the interested users via a gateway, often using the internet or any other communication path \cite{G9}. It should be noted that with the development of machine-to-machine communications, it is possible to have sensors and machines directly connected to cellular network based mobile Internet.

At the time of writing, the most common way of constructing WSNs relies on the ZigBee communications protocol, which complies with the IEEE 802.15.4 standard, outlining the specifications of both the physical layer (\gls{PHY}) and the medium access control (\gls{MAC}) layer. This is widely regarded as the \textit{de facto} standard for WSNs \cite{Khanafer2014}.
A WSN operates in the unlicensed industrial, scientific, and medical (\gls{ISM}) band, in which it coexists with many other successful communication systems, such as the IEEE 802.11 standard based wireless local area network (\gls{WLAN}) and the 802.15.1 standard based Bluetooth communication systems. Therefore, a WSN may face the challenge of co-channel interferences imposed by both other WSNs and other co-existing heterogeneous wireless systems. This coexistence problem may substantially affect the performance of WSNs.

\subsection{Applications}

\begin{figure*}[htb]
\centering
\includegraphics[width=15cm]{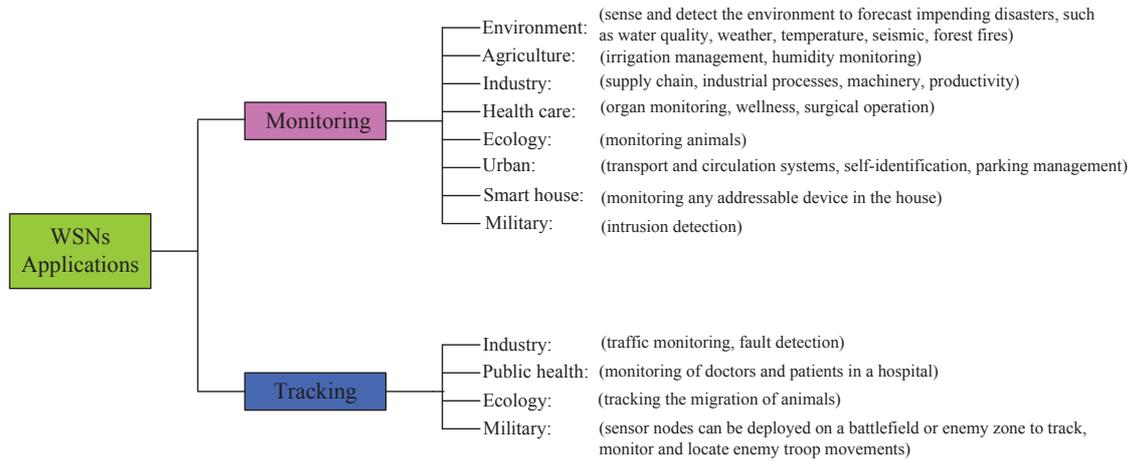}
\caption{Taxonomy of WSN applications.}
\label{sys3}
\end{figure*}

In WSNs, sensor nodes are generally deployed randomly in the majority of application domains. When the sensor nodes are deployed in hostile remote environments, they may be equipped with high-efficiency energy harvesting devices (e.g. solar cells) for extending the network lifetime\cite{energy_harvesting_WSN}. Numerous practical applications of WSNs have been rolled out with the advancement of technologies. In general, the applications of WSNs can be classified into two types: monitoring and tracking. Monitoring is used for analyzing, supervising and carefully controlling operation of a system in real-time. Tracking is generally used for following the change of an event, a person, an animal, and so on.
Existing monitoring applications include indoor/outdoor environmental monitoring \cite{Arampatzis2005},
industrial monitoring \cite{Somappa2014}, precision agriculture (e.g., irrigation management and crop disease prediction) \cite{Zhu2011}, biomedical or health monitoring \cite{Milenkovi2006}, electrical network monitoring \cite{Isaac2011}, military location monitoring \cite{Boukerche2008}, and so forth. Tracking applications include habitat tracking \cite{Prachi2015}, traffic tracking \cite{Rashid2016}, military target tracking \cite{Tafa2012}, etc.
We summarize their classification in Fig. \ref{sys3}.
\begin{figure*}[t]
\centering
\includegraphics[width=16cm]{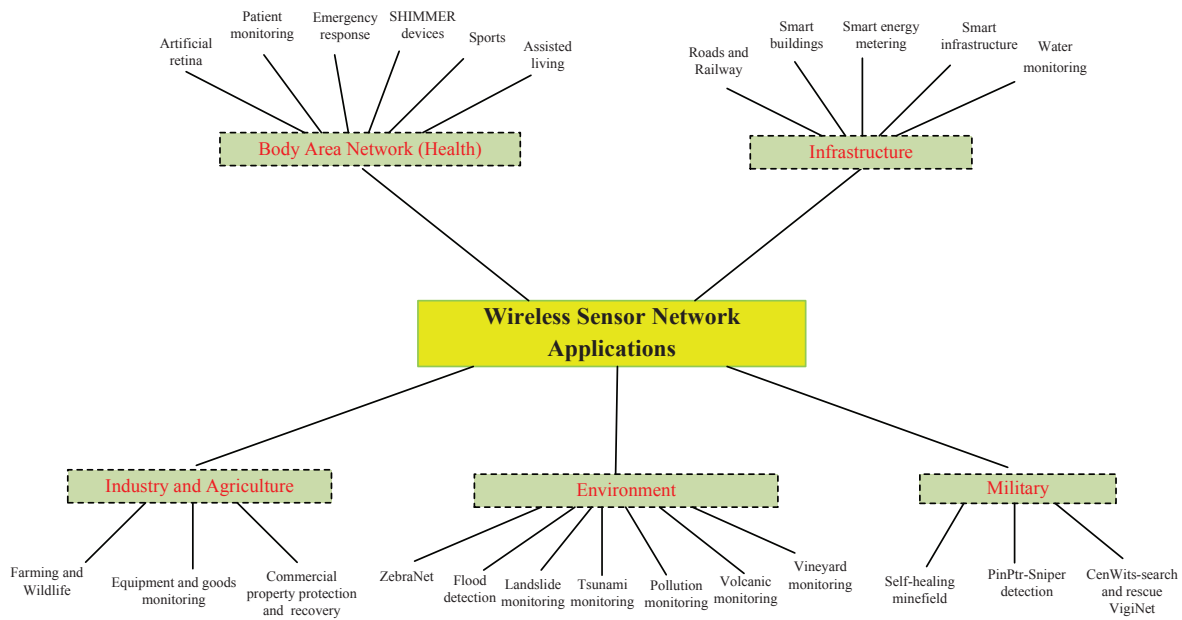}
\caption{Overview of WSN applications in real environments. Here, SHIMMER represents the Intel digital health group's ``sensing health with intelligence, modularity, mobility, and experimental re-usability'' \cite{G9}.}
\label{sys4}
\end{figure*}

A more detailed portrayal of WSN applications is given in Fig. \ref{sys4}. For instance, in environmental monitoring, WSNs can help us perform forest fire detection, flood detection, forecast of earthquakes and eruptions, pollution monitoring, etc. \cite{Arampatzis2005,Spachos_SensorsJ2016}.
In industry and agriculture, WSNs can sense and detect farming and wildlife, monitor equipment and goods, protect commercial property, predict crop disease and production quality, control pests and diseases, etc. \cite{Somappa2014}, \cite{Zhu2011}.
In healthcare and biomedical applications, WSNs can be utilized for diagnostics, distance-monitoring of patients and their physiological data, as well as for tracking the medicine particles inside patients' body, etc. \cite{Milenkovi2006}. In particular, WSN based wireless body area networks (\glspl{WBAN}) can help monitor human body functions and characteristics (e.g., artificial retina, vital signs, automatic drug delivery, etc.), acting as an \textit{in vitro} or \textit{in vivo} diagnostic system \cite{Movassaghi2014}. In the infrastructure, WSNs have been widely used for monitoring the railway systems and their components, such as bridges, rail tracks, track beds, track equipment, as well as chassis, wheels, and wagons that are closely related to rolling stock quality\cite{Chinrungrueng2006}.
WSNs also allow users to manage various appliances both locally and remotely for building automation applications\cite{Alam_Sensors2014}.
In military target tracking and surveillance, a WSN can assist in intrusion detection and identification. Specific examples include enemy troop and tank movements, battle damage assessment, detection and reconnaissance of biological, chemical and nuclear attacks, etc. \cite{Tafa2012}.

Furthermore, several novel WSN application scenarios, such as the Internet of Things \cite{Mainetti2011}, cyber-physical systems \cite{Wu2011} and smart grids \cite{Fadela2015}, among others, have adopted new design approaches that support multiple concurrent applications on the same WSN.
The applications of WSNs are not limited to the areas mentioned in this paper. The future prospects of WSN applications are promising in terms of revolutionizing our daily lives.

\subsection{MOO Metrics}

In this subsection, a succinct overview of the most popular optimization objectives of WSNs is provided.
Over the past years, a number of research contributions have addressed diverse aspects of WSNs, including their protocols\cite{Q3}, routing\cite{Q4}, energy conservation\cite{G1}, lifetime\cite{Q5} and so forth. The QoS, as perceived by the users or applications, was given insufficient attention at the beginning. However, at the time of writing, how to provide the desired QoS is becoming an increasingly important topic for researchers.  Different applications may have their own specific QoS requirements, but some of the more commonly used metrics for characterizing QoS are the coverage area and quality, the delay, the number of active nodes, the bit-error rate (\gls{BER}) and the overall WSN lifetime.

\begin{figure*}[t]
\centering
\includegraphics[width=12cm]{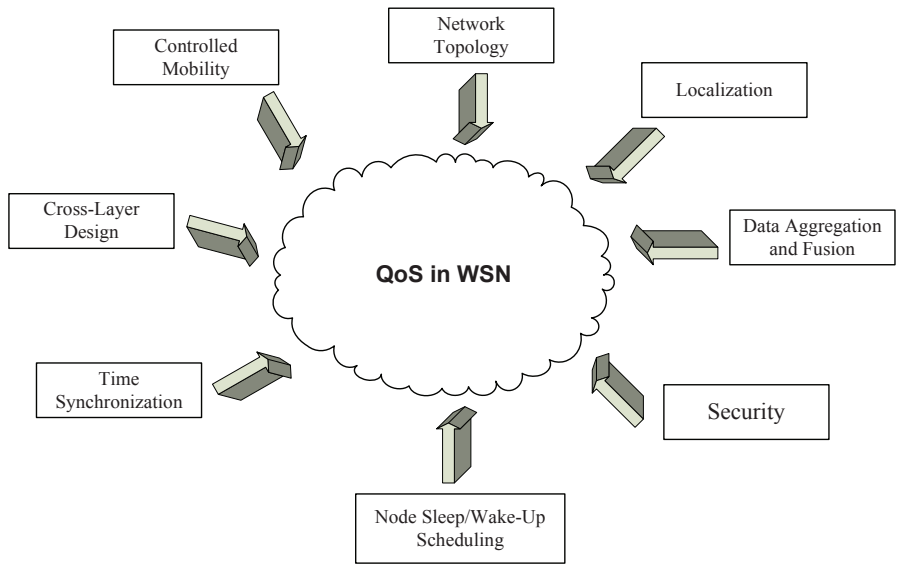}
\caption{The interplay of factors affecting the QoS in WSNs.}
\label{sys5}
\end{figure*}

There are many other QoS metrics worth mentioning, and a range of factors affecting the QoS in WSNs are portrayed in Fig. \ref{sys5}, which was directly inspired by \cite{S1} and reflects the application requirements of a WSN. It is indeed plausible that the network's performance can be quantified in terms of its energy conservation, lifetime, and QoS-based metrics in specific applications.
However, multiple metrics usually conflict with each other. For example, when more energy is consumed by the nodes, the operating lifetime of the network reduces. Similarly, if more active nodes are deployed in a given region, a lower per-node power is sufficient for maintaining connectivity, but the overall delay is likely to be increased due to the increased number of hops. Hence, an application-specific compromise has to be struck between having more short hops imposing a lower power dissipation but higher delay and having more longer hops, which reduces the delay but may increase the transmit-power dissipation.
\subsubsection{Coverage}
``Coverage'' is one of the most important performance metrics for a sensor network. In other wireless communication networks, coverage typically means the radio coverage. By contrast, coverage in the context of WSNs corresponds to the sensing range, while connectivity corresponds to the communication range. WSN coverage can be classified into three types: \textit{area coverage}, \textit{point coverage} and \textit{barrier coverage} \cite{T41}.  In \textit{area coverage}, the coverage quality of an entire two-dimensional (\gls{2D}) region is considered, where each point in the region is observed by at least one sensor node. In \textit{point coverage}, the objective is to simply guarantee that a finite set of points in the region are observed by at least one sensor node. \textit{Barrier coverage} usually deals with the detection of movement across a barrier of sensor nodes. The most richly studied coverage problem in the WSN literature is the area coverage problem. The characterization of the coverage varies depending both on the underlying models of each node's field of view and on the metric used for appraising the collective coverage. Several coverage models \cite{Meguerdichian_2001:coverage_WSN,Meguerdichian_2001:Exposure_WSN, Huang_2005:coverage_models} have been proposed for different application scenarios. A coverage model is normally defined with respect to the sensing range of a sensor node. The most commonly used node coverage model is the so-called \textit{sensing disk model}, where all points within a disk centered at the node are considered to be covered by the node \cite{M41}. More specifically, a point $p$ is regarded to be covered/monitored by at least a node $v$ if their Euclidean distance is less than the sensing range $R_s$ of node $v$.

Given a set of nodes, finding the optimal positions of these nodes to achieve maximum coverage is in general an NP-complete problem\cite{M2}. There are many different ways of solving the coverage optimization problem suboptimally. Here, in order to make the coverage problem more computationally manageable, we consider an area $A$ represented by a rectangular grid, which is divided into $G = xy$ rectangular cells of identical size, and let $(x_j, y_j)$ denote the coordinate of node $j$.
As a result, the network coverage $C_v(x)$ is defined as the percentage of the adequately covered cells over the total cells of $A$ and it is evaluated as follows \cite{C9,M2}:

\begin{align}
    C_v(x)=\frac{\sum_{x'=0}^x\sum_{y'=0}^y g(x',y')}{xy},
\end{align}
and
\begin{small}
\begin{eqnarray}
g(x',y')=
\begin{cases}
    1, ~ \mathrm{if}~ \exists~j\in\{1,\ldots,N\},d_{(x_j,y_j),(x',y')}\leq R_s,\\
    0, ~ \mathrm{otherwise},
\end{cases}
\end{eqnarray}
\end{small}where $N$ is the number of nodes, $g(x',y')$ is the monitoring status of the cell centered at $(x',y')$, $R_s$ is the sensing range of a node, and $d_{(x_j,y_j),(x',y')}$ is the distance from the location of node $j$ to the cell centered at $(x',y')$. Having a better coverage also leads to a higher probability of detecting the event monitored\cite{Alsheikh_Surveys2015}.

\subsubsection{Network Connectivity}

Another issue in WSN design is the network connectivity that is dependent on the selected communication protocol \cite{D2}. Two sensor nodes are directly connected if the distance of the two nodes is smaller than the communication range $R_c$. Connectivity only requires that the \textit{location of any
active node} is within the communication range of one or more active nodes,
so that \textit{all} active nodes can form a connected communication network. The most common protocol relies on a cluster-based architecture, where all nodes in the same cluster can directly communicate with each other via a single hop and all nodes in the same cluster can communicate with all nodes in the neighboring clusters via the cluster head. In a given cluster only a single node acts as the cluster head, which has to be active in terms of collecting all information gleaned by the other nodes for the sake of maintaining connectivity. In cluster-based WSNs, the connectivity issues tend to hinge on the number of nodes in each cluster (because a cluster head can only handle up to a specific number of connected nodes), as well as on the coverage issues related to the ability of any location to be covered by at least one active sensor node.

For an area represented by a rectangular grid of size $x\times y$, let $R_{c_i}$ and $R_{s_i}$ denote the communication range and sensing range of the $i$th sensor node, respectively. To guarantee each sensor node is placed within the communication range of at least another sensor node and to prevent sensor nodes from becoming too close to each others, the objective function associated with the network connectivity can be expressed as \cite{Syarif2014}
\begin{align}
    f_{con}=\sum_{i=1}^{x\times y}1-e^{-(R_{c_i}-R_{s_i})},
\end{align}
where $R_{c_i}-R_{s_i}>0$ has to be satisfied for achieving network connectivity.

Maintaining the network's connectivity is essential for ensuring that the messages are indeed propagated to the appropriate sink node or base station,  and the loss of connectivity is often treated as the end of the network's lifetime. Network connectivity is closely related to the coverage and energy efficiency of WSNs. To elaborate a little further, substantial energy savings can be achieved by dynamic management of node duty cycles in WSNs having high node density. In this method, some nodes can be scheduled to sleep (or enter a power-saving mode), while the remaining active nodes provide continuous service. As far as this approach is concerned, a fundamental problem is to minimize the number of nodes that remain active while
still achieving acceptable QoS. In particular, maintaining an adequate sensing coverage and network connectivity with the active
nodes is a critical requirement in WSNs. The relationship between coverage and connectivity hinges on the ratio of the communication range to the sensing range. A connected network may not be capable of guaranteeing adequate coverage regardless of
the ranges. By contrast, in\cite{Wang_2003:coverage_vs_connectivity,Xing_2005:coverage_vesus_connectivity} the authors presented a sufficient condition for guaranteeing network connectivity, which states that for a set of nodes that cover a convex region, the network remains connected if $R_c \ge 2R_s$. There exist tighter relationships between $R_c$ and $R_s$ for achieving
network connectivity, provided that adequate sensing coverage is
guaranteed\cite{Ammari_2006:integrated_coverage_connectivity,Ammari_2008:integrated_coverage_connectivity,
Li_2010:connectivity_coverage_relation}. Intuitively, if the communications range of sensor nodes is sufficiently large, then maintaining connectivity is not a problem, because in this case there always exists a node to communicate with. A more in-depth discussion of the relationship between coverage, connectivity and energy efficiency of WSNs can be found in \cite{Wang_2003:coverage_vs_connectivity,Xing_2005:coverage_vesus_connectivity, Ammari_2006:integrated_coverage_connectivity, Ammari_2008:integrated_coverage_connectivity,Li_2010:connectivity_coverage_relation}.

\subsubsection{Network Lifetime}
Another important performance metric in WSNs is their lifetime. Tremendous research efforts have been invested into solving the problem of prolonging network lifetime by energy conservation in WSNs. Indeed, the energy source of each node is generally limited, while recharging or replacing the battery at the sensors may be impossible. Hence, both the radio transceiver unit and the sensor unit of each node have to be energy-efficient, and it is vitally important to maximize the attainable network lifetime \cite{A1}, defined as the time interval between the initialization of the network and the depletion of the battery of any of the sensor nodes.

For the simplicity of exposition, typically all sensor nodes are assumed to be of equal importance, which is a reasonable assumption, since the ``death'' of one sensor node may result in the network becoming partitioned, or some area requiring monitoring to be uncovered. Thus, the network's lifetime is defined as the time duration from the application's first activation to the time instant when any of the sensor nodes in the cluster fails due to its depleted energy source.

More explicitly, this objective can be formulated as
\begin{align}
    T_{net}=\min T_j,
\end{align}
where $j = 1, 2, \cdots, N$.

The lifetime of a sensor node is generally inversely proportional both to the average rate of its own information generated and to the information relayed by this node. Hence, the network's lifetime is also partially determined by the source rates of all the sensor nodes in the network.

\subsubsection{Energy Consumption}

Sensor nodes are equipped with limited battery power and the total energy consumption of the WSN is a critical consideration. Each node consumes some energy during its data acquisition, processing and transmission phases. For instance, in a heterogeneous WSN, different sensor nodes might have diverse power and data processing capabilities. Hence, the energy consumption of a WSN depends both on the Shannon capacity of the channels among the nodes and on these nodes' functionality. The energy consumption of a path $P$ is the sum of the energy expended at each node along the path,
hence the total energy consumption $E(P)$ of a given path is given by \cite{R4}
\begin{align}
    E(P)=\sum_{i=0}^L(t_{i}^a+t_{i}^p)\times P_i^o + P_{i}^t\times t^m,
\end{align}
where $t_{i}^a$ and $t_{i}^p$ indicate the time durations of data acquisition and data processing taking place at node $i$, respectively. $L$ is the number of nodes on the given path. Furthermore, $t^m$ is the message transmission
time, while $P_i^o$ and $P_{i}^t$ denote the operational power and transmission power dissipation of node $i$, respectively.

\subsubsection{Energy Efficiency}

Energy efficiency is a key concern in WSNs, and this metric is closely related to network lifetime in the particular context of WSNs\footnote{Note that the concept of energy efficiency is also widely used in green communications \cite{Q6,Q7,Q8,Q9,Taufik_2016:delay_EE}. The definition of energy efficiency has several variants. It is typically defined as the ratio of the spectral efficiency (bits/second/Hz) to the power dissipation of the system considered. Hence, its unit is bits/second/Hz/Watt or equivalently bits/Hz/Joule. Alternatively, it can be defined as the power-normalized transmission rate, and hence its unit becomes bits/second/Watt or bits/Joule.}\cite{Mao_Parallel2011}.
As an example, herein the energy efficiency of node $i$ is defined as the ratio of the transmission rate to the power dissipation. Explicitly, it is formulated as
\begin{align}
    \eta_i=\frac{W\log_2(1+\gamma_i)}{p_i},
\end{align}
where $W$ denotes the communication bandwidth, $p_i$ is the transmission power of node $i$ and $\gamma_i$ denotes the signal-to-interference-plus-noise ratio (\gls{SINR}) at the destination receiver relative to node $i$, respectively.

Due to the limited energy resources of each node, we have to utilize these nodes in an efficient manner so as to increase the lifetime of the network \cite{D44}. There are at least two approaches to deal with the energy conservation problem in WSNs. The first approach is to plan a schedule of active nodes while enabling the other nodes to enter a sleep mode. The second approach is to dynamically adjust the sensing range of nodes for the sake of energy conservation.

\subsubsection{Network Latency}

For a WSN, typically a fixed bandwidth is available for data transfer between nodes. Again, having an increased number of nodes results in more paths becoming available for simultaneously routing packets to their destinations, which is beneficial for reducing the latency. Meanwhile, this may also degrade the latency that increases proportionally to the number of nodes on the invoked paths. This is due to additional contention for the wireless channel when the node density increases, as well as owing to routing and buffering delays.

The delay between source node $u_{so}$ and sink node $u_{si}$, denoted as $D_{u_{so},u_{si}}$, is defined as the time elapsed between the departure of a collected data packet from $u_{so}$ and its arrival to $u_{si}$, and is given by\cite{C24, Ammari_2005:tradeoff_delay_energy}
\begin{align}
    D_{u_{so},u_{si}}&= (T_q+T_p+T_d)\times N(u_{so},u_{si}) \nonumber \\
                       &= c \times N(u_{so},u_{si}) \nonumber \\
                       &\propto  N(u_{so},u_{si}),
\end{align}
where $T_q$ is the queue delay per intermediate forwarder, $T_p$ is the propagation delay and $T_d$ is the transmission delay. All of them are, for the sake of simplicity, regarded as constants and collectively denoted by $c$. Finally, $N(u_{so},u_{si})$ denotes the total number of data disseminators between $u_{so}$ and $u_{si}$.
As a consequence, the minimization of the delay corresponds to minimizing the number of intermediate forwarders between the source and the sink. It is worth noting that Haenggi $et~al.$ \cite{C1} astutely argued that long-hop based routing is a very competitive strategy compared to short-hop aided routing in terms of latency, albeit this design dilemma also has ramifications as to the scarce energy resource of the nodes.

\subsubsection{Differentiated Detection Levels}

Differentiated sensor network deployment is also an important issue. In many real WSN applications, such as underwater sensor deployments or surveillance applications, certain parts of the supervised region may require extremely high detection probabilities if these parts constitute safety-critical geographic area. However, in the less sensitive geographic area, relatively low detection probabilities have to be maintained for reducing the number of nodes deployed, which corresponds to  reducing the cost imposed. Therefore, different geographic areas require different densities of deployed nodes, and the sensing requirements are not necessarily uniformly distributed within the entire supervised region.

Let us use $d((m,n),(i,j))$ to denote the Euclidean distance between the coordinates $(m,n)$ and $(i,j)$.
A probabilistic node detection model can be formulated as\cite{D14,Zou_2004:uncertainty,Zou_2004:Sensor_deployment}
\begin{small}
\begin{eqnarray}
p((m,n),(i,j))=
\begin{cases}
    e^{-ad((m,n),(i,j))}, ~ d((m,n),(i,j))\leq R_s, \\
    0, ~~~~~~~~~~~~~~~~~~ d((m,n),(i,j))> R_s,
\end{cases}
\end{eqnarray}
\end{small}where $a$ is a parameter associated with the physical characteristics of the sensing device and $R_s$ is the sensing range.

\subsubsection{Number of Nodes}

Each sensor node imposes a certain cost, including its production, deployment and maintenance. As a result, the total cost of the WSN increases with the number of sensor nodes. When deploying a WSN in a battleground, sensor nodes have to operate as stealthily as possible to avoid being detected by the enemy. This implies that the number of nodes has to be kept at a minimum in order to reduce the probability of any of them being discovered. \cite{E1,F1,E2,E3,D88,C13} are dedicated to optimal node deployment by considering the accomplishment of the specified goals at a minimum cost.

Minimizing the number of active nodes is equivalent to maximizing the following objective \cite{Q11}:
\begin{align}
    f(K')=1-\frac{|K'|}{|K|},
\end{align}
where $|K'|$ is the number of active nodes and $|K|$ is the total number of nodes.

\subsubsection{Fault Tolerance}

Sensor nodes may fail, for example, due to the surrounding physical conditions or when their energy runs out. It may be difficult to replace the existing nodes, hence the network has to be fault tolerant in order to prevent individual failures from reducing the network lifetime \cite{R1,Bhuiyan_TComputers2015}. In other words, fault tolerance can be viewed as an ability to maintain the network's operation without interruption in the case of a node failure, and it is typically implemented in the routing and transport protocols.
The fault tolerance or reliability $R_{k}(t)$
of a sensor node can be modeled using the Poisson distribution in order to capture the
probability of not having a failure within the time interval $(0,t)$ as\cite{A11}:
\begin{equation}
R_{k}(t) = e^{-\lambda_k t},
\end{equation}
where  $\lambda_k$ is the failure rate of sensor node $k$ and
$t$ is the time period.

Numerous studies have been focused on forming $k$-connected WSNs\cite{M31, Wang_2003:coverage_vs_connectivity,Xing_2005:coverage_vesus_connectivity}. The $k$-connectivity implies that there are $k$ independent paths in the full set of the pair of nodes. For $k\geq 2$, the network can tolerate some node and link failures. Due to the many-to-one interaction pattern, $k$-connectivity is a particularly important design factor in the neighborhood of base stations and guarantees maintaining a certain communication capacity among the nodes \cite{M31}.

\subsubsection{Fair Rate Allocation}

It is important to guarantee that the sink node receives information from all sensor nodes in a fair manner when the bandwidth is limited.
The accuracy of the received source information depends on the allocated source rate. Simply maximizing the total throughput of the network is insufficient for guaranteeing the specific application's performance, since this objective may only be achieved at the expense of sacrificing the source rate supposed to be allocated to some nodes\cite{B1}. For example, in a sensor network that tracks the mobility of certain objects in a large field of observation, lower rates impose a reduced location tracking accuracy and vice versa. By simply maximizing the total throughput instead of additionally considering the above fairness issues among sources, we may end up with a solution that shuts off many sources in the network and enables only those sources whose transport energy-cost to the sink is the lowest. Hence, considering the fairness of rate allocation among different sensor nodes is of high significance.

An attractive methodology of achieving this goal is to adopt a network utility maximization (\gls{NUM}) framework \cite{B1}, in which a concave, non-decreasing and twice differentiable utility function $U_i(x_i)$ quantifies the grade of satisfaction of sensor node $i$ with the assigned rate $x_i$, and the goal is to maximize the sum of individual utilities. A specific class of utility functions that has been extensively used for achieving fair resource allocation in economics and distributed computing\cite{Q18} is formulated as:

\begin{eqnarray}
U^\alpha({\bf x})=
\begin{cases}
    \log {\bf x} & \alpha=1, \\
    \frac{1}{1-\alpha}{\bf x}^{1-\alpha} & \alpha > 1,
\end{cases}
\end{eqnarray}
where ${\bf x} = (x_i, \forall i)$ and the functional operations are elementwise.
When we have $\alpha=1$, the above utility function leads to the so-called proportional fairness, whereas when $\alpha\rightarrow \infty$, this utility function leads to max--min fairness\footnote{The \textit{max-min criterion} constitutes one of the most commonly used fairness metrics \cite{Q18}, in which a feasible flow rate vector ${\bf x} = (x_i, \forall i)$ can be interpreted as being max-min fair if the rate $x_i$ cannot be increased without decreasing some $x_j$ that is smaller than or equal to $x_i$, $\forall i \neq j$.
The concept of \textit{proportional fairness} was proposed by Kelly \cite{Q19}. A vector of rates ${\bf x}^* = (x_i^*, \forall i)$
is proportionally fair if it is feasible (that is, $x^*\geq 0$ and ${\bf A}^T{\bf x}^*\leq {\bf c}$) and if for any other feasible vector ${\bf x} = (x_i, \forall i)$, the aggregate of proportional change is non-positive, i.e., $\sum_i\frac{x_i-x_i^*}{x_i^*}\leq 0$. Here ${\bf c} = (c_m, \forall m)$ with each element denoting the source rates to be allocated. $\mathbf{A}=(A_{im}, \forall i,m)$ is a matrix that satisfies: if node $i$ is allocated source rate $m$, $A_{im}=1$, otherwise $A_{im}=0$.}.

\subsubsection{Detection Accuracy}

Having a high target detection accuracy is also an important design goal for the sake of achieving accurate inference about the target in WSNs.
Target detection accuracy is directly related to the timely delivery of the density and latency information of the WSN. Assume that a node $k$ receives a certain amount of energy $e_k(u)$ from a target located at location $u$ and $K_o$ is the energy emitted by the target. Then, the signal energy $e_k(u)$ measured by node $k$ is given by \cite{R4}
\begin{align}
e_k(u)=K_o/(1+\alpha d_k^p),
\end{align}
where $d_k$ is the Euclidean distance between the target location and the location of node $k$, $p$ is the pathloss exponent that typically assumes values in the range of $[2,4]$, while $\alpha$ is an adjustable constant.

\subsubsection{Network Security}

\begin{figure}[tbp]
\centering
\includegraphics[width=2.5in]{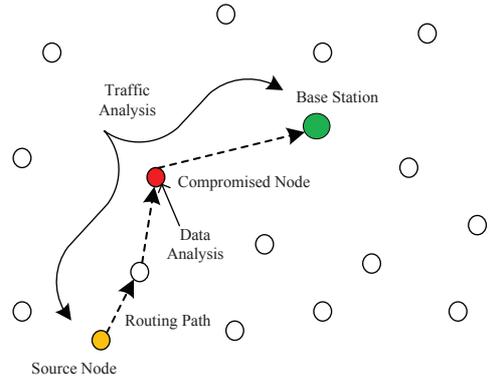}
\caption{Two types of privacy attacks in a WSN \cite{G6}: Data analysis attack and traffic analysis attack conducted by a malicious node.}
\label{sys11}
\end{figure}

Sensor nodes may be deployed in an uncontrollable environment,
such as a battlefield, where an adversary might aim for launching physical attacks in order to capture sensor nodes or to deploy counterfeit ones. As a result, an adversary may retrieve private keys used for secure communications by eavesdropping and decrypt the
communications of the legitimate sensors. Recently, much attention has been paid to the security of WSNs. There are
two main types of privacy concerns, namely data-oriented and context-oriented concerns \cite{G6}. Data-oriented concerns focus on the privacy of data collected from a WSN, while context-oriented concerns concentrate on contextual information, such as the location and timing of traffic flows in a WSN. A simple illustration of the two types of security attacks is depicted in Fig. \ref{sys11}.

We can observe from Fig. \ref{sys11} that a malicious node of the WSN abuses its ability of decrypting data in order to compromise the payload being transmitted in the case of data analysis attack. In traffic-analysis
attacks, the adversary does not have the ability to decrypt data payloads. Instead, it eavesdrops to intercept the transmitted data and tracks the traffic flow on a hop-by-hop basis.

For the sake of improving the network security, we can minimize the loss of privacy that is calculated based on information theory as \cite{F23}
\begin{align}
\zeta=1-2^{-I(S,X)},
\end{align}
where $I(S,X)$ is the mutual information between the random variables $S$ and $X$. More specifically, $S$ represents the current position of the node of interest, $X$ is the observed variable known to the attacker and correlated to $S$, while $I(S,X)=H(S)-H(S|X)$ with $H(\cdot)$ denoting the entropy. Additionally, we can also minimize the probability of eavesdropping in a WSN, as presented in\cite{F30}, to improve the network security.

Table \ref{existingapproaches} summarizes the representative existing contributions to optimizing the particular WSN performance metrics mentioned above.

%\multirow{2}*{\backslashbox{Metrics}{Approaches}}

\begin{table*}\scriptsize
\centering
\setlength{\tabcolsep}{2pt}
\renewcommand{\arraystretch}{1.3}
\extrarowheight 3pt
\caption{Major Existing Approaches for Evaluating/Improving/Optimizing Each Metric.}
\begin{tabular}{| c | c | p{2.5cm} | c | c | c | c | c | c |c|}
%\toprule
\hline
&&\multirow{2}{*}{References} & \multicolumn{7}{|c|}{\textbf{Major Evaluation/Improvement/Optimization Approaches}} \\
\cline{4-10}
 & & & Protocol design & Mathematical programming & EAs & SIOAs & Hybrid algorithms & Theoretical analysis & Simulator \\
\hline
\multirow{3}{*}{\begin{sideways}{Coverage}\end{sideways}} & Area coverage& \cite{M2,M1,Yun_TNET2010,Li_IET2010,D44,E4,Abo-Zahhad2016} &   &\checkmark &\checkmark &  & \checkmark& &\checkmark \\
\cline{2-10}
 & Point coverage&\cite{Zhao_TNET2008,Shih_JNCA2009} & & & &&  &  &\checkmark \\
\cline{2-10}
& Barrier coverage & \cite{Ram_TMC2007,Chen_TMC2010,Sakai_Sensors2015}&   &\checkmark& &  &  & &\checkmark \\
%\cline{1-10}
\hline
\multicolumn{2}{|c|}{Network connectivity} & \cite{Ma_Sensors2016,Zhang_JSAC2010,E4}& & & & &  &  & \checkmark \\
\hline
\multicolumn{2}{|c|}{Network lifetime} & \cite{M2,M1,Zhao_TPDS2012,Shah-Mansouri_TWC2010,Rout_TWC2013,Wang_Sensors2014,E4,C9,B1,C6,C7,D0,C10}
& \checkmark &\checkmark & \checkmark&  & \checkmark & \checkmark & \checkmark \\
\hline
\multicolumn{2}{|c|}{Energy consumption} & \cite{Miller_Mobile2005,Baek_JSAC2004,Elhoseny_CL2015,Abo-Zahhad2016,D22,D11,F2,C2,Yao2015TNet,C21,C24,D55,F22,F21,C5,C4,F211,F24}& \checkmark & \checkmark& \checkmark & \checkmark  & \checkmark & \checkmark&\checkmark \\
\hline
\multicolumn{2}{|c|}{Energy efficiency} & \cite{D44,Gao_TVT2010,Fang_Electronics2010,Muruganathan_Magazine2005,X1,Gutierrez2015,D66}  &\checkmark& & & \checkmark & \checkmark & & \checkmark \\
\hline
\multicolumn{2}{|c|}{Network latency} & \cite{Sahoo_Mobile2010,Cheng_Sensors2011,D44,D22,D11,F2,C2,Yao2015TNet,C21,C24,D55,F22,F21,C5,C4,F211,F24} & \checkmark& \checkmark &\checkmark & \checkmark & \checkmark & \checkmark& \checkmark \\
\hline
\multicolumn{2}{|c|}{Differentiated detection levels} &\cite{D14,Aitsaadi_VTC2008} &  & \checkmark&  &   & & &\checkmark \\
\hline
\multicolumn{2}{|c|}{Number of nodes} & \cite{Wang_IJDSN2014,E1,F1,E2,E3,D88,C13,Enayatifar2014,F212}&  & &\checkmark & \checkmark&  & & \checkmark\\
\hline
\multicolumn{2}{|c|}{Fault tolerance} &\cite{Han_Mobile2010,WangJSAC2005,Bhuiyan_TComputers2015} &   && & &   & & \checkmark \\
\hline
\multicolumn{2}{|c|}{Fair rate allocation} & \cite{B1,Yao_ICC2013,Narayanan_Workshops2011,C7} &  & \checkmark& &  &   & & \checkmark \\
\hline
\multicolumn{2}{|c|}{Detection accuracy} & \cite{R4,Lee_CC2008,Tan_TWC2010,Yan_TWC2014,C14,C15}& \checkmark& & \checkmark& &  & \checkmark
& \checkmark\\
\hline
\multicolumn{2}{|c|}{Network security} & \cite{Ren_Mobile2008,Chen_Surveys2009,F23,F30} &\checkmark & & & &  &  \checkmark & \checkmark \\
\hline
\end{tabular}
\label{existingapproaches}
\end{table*}

\section{Techniques of MOO \label{sec5}}

In this section, we briefly present the MOO techniques proposed in the literature for tackling various important problems in WSNs.
\begin{figure*}[t]
\centering
\includegraphics[width=13cm]{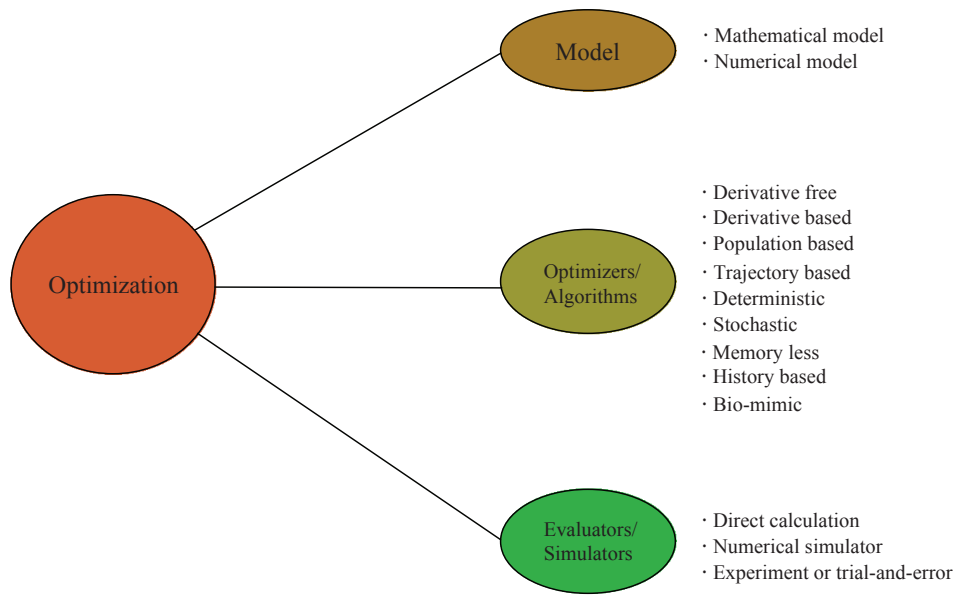}
\caption{A simple illustration of optimization process.}
\label{sys6}
\end{figure*}

\subsection{Optimization Strategies}

Optimization covers almost all aspects of human life and work. In practice, the resources are limited, hence optimization is important. Most research activities in computer science and engineering involve a certain amount of modeling, data analysis, computer simulations and mathematical optimization. This branch of applied science aims for finding the particular values of associated variables, which results in either the minimum or the maximum values of a single objective function or multiple objective functions \cite{B11}.  A typical optimization process is composed of three components \cite{Jabbar2013}: the model, the optimizer/algorithm and the evaluator/simulator, as shown in Fig. \ref{sys6}. The representation of the physical problem is carried out by using mathematical formulations to establish a mathematical model.

As an important step of solving any optimization problem, an efficient optimizer or algorithm has to be designed for ensuring that the optimal solution is obtained. There is no single algorithm that is suitable for all problems.
Optimization algorithms can be classified in many ways, depending on the specific characteristics that we set out to compare. In general, optimization algorithms can be classified as:
\begin{enumerate}
\item Finitely terminating algorithms, such as the family of simplex algorithms and their extensions, as well as the family of combinatorial algorithms;
\item Convergent iterative methods that
\begin{itemize}
\item evaluate Hessians (or approximate Hessians, using finite differences), such as Newton's method and sequential quadratic programming;
\item evaluate gradients or approximate gradients using finite differences (or even subgradients), such as quasi-Newton methods, conjugate gradient methods, interior point methods, gradient descent (alternatively, "steepest descent" or "steepest ascent") methods, subgradient methods, bundle method of descent, ellipsoid method, reduced gradient method, and simultaneous perturbation based stochastic approximation methods;
\item and evaluate only function values, such as interpolation methods and pattern search methods.
\end{itemize}
\item Heuristics/metaheuristics that can provide approximate solutions to some optimization  problems.
\end{enumerate} Recently, bio-mimetic heuristics/metaheuristics based strategies have been widely used for solving MOPs, since they are capable of obtaining near-optimal solutions to optimization problems characterized by non-differential nonlinear objective functions, which are particularly hard to deal with using classical gradient- or Hessian-based algorithms.

A general MOP consists of a number of objectives to be simultaneously optimized and it is associated with a number of inequality and equality constraints.
Without loss of generality, a multi-objective minimization problem having $n$ variables and $m$ ($m>1$) objectives can be formulated as
\begin{align}
    \min f({\bf x})=&\min[f_1({\bf x}),f_2({\bf x}),\cdots,f_m({\bf x})]  \nonumber\\
\mbox{s.t.}\quad
   &g_i({\bf x})\leq 0,~i=1,2,\ldots,m_{ie},  \nonumber\\
   &h_j({\bf x})=0,~j=1,2,\ldots,m_{eq},
\label{c}
\end{align}
where we have ${\bf x}\in \mathbf{R}^n$ with $\mathbf{R}^n$ being the decision space, and $f({\bf x})\in \mathbf{R}^m$ with $\mathbf{R}^m$ representing the objective space. The objective functions of (\ref{c}) are typically in conflict with each other in the real world. Explicitly, the improvement of one of the objectives may result in the degradation of other objectives, thus it is important to achieve the Pareto-optimality, which represents the conditions when none of the objective functions can be reduced without increasing at least one of the other objective functions \cite{T1}. For the minimization of $m$ objectives $f_1({\bf x}),f_2({\bf x}),\cdots,f_m({\bf x})$, we have the following definitions.
\begin{itemize}
 \item
\emph{Non-dominated solutions}: A solution $\bf a$ is said to dominate a solution $\bf b$ if and only if \cite{L1}:

$(1)~f_i({\bf a})\leq f_i({\bf b})~~\forall i\in\{1,2,\cdots,m\}$,

$(2)~f_i({\bf a})< f_i({\bf b})~~\exists i\in\{1,2,\cdots,m\}$.

Solutions that dominate the others but do not dominate themselves are termed non-dominated solutions.

\item
\emph{Local optimality in the Pareto sense}: A solution $\bf a$ is said to be locally optimal in the Pareto sense, if there exists a real $\epsilon > 0$ such that there is no other solution $\bf b$ dominating the solution $\bf a$ with ${\bf b} \in\mathbf{R}^n\cap B({\bf a},\epsilon)$, where $B({\bf a},\epsilon)$ shows a bowl having a center $\bf a$ and a radius $\epsilon$.

\item
\emph{Global optimality in the Pareto sense}: A solution $\bf a$ is globally optimal in the Pareto sense, if there does not exist any vector $\bf b$ that dominates the vector $\bf a$. The main difference between global and local optimality lies in the fact that for global optimality we no longer have a restriction imposed on the decision space $\mathbf{R}^n$ .

\item
\emph{Pareto-optimality}: A feasible solution is said to be Pareto-optimal, when it is not dominated by any other solutions in the feasible space. PS, which is also often referred to as the efficient set, is the collection of all Pareto-optimal solutions and their corresponding images in the objective space are termed the PF.
\end{itemize}
\begin{figure}[t]
\begin{centering}
\subfloat[PF of unconstrained MOP]{\begin{centering}
\includegraphics[width=6.0cm]{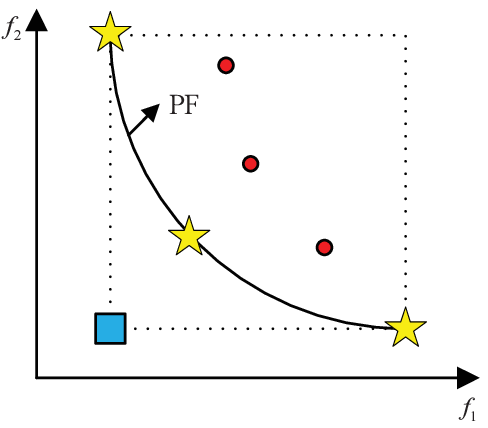}
\par\end{centering}}

\subfloat[PF of constrained MOP]{\begin{centering}
\includegraphics[width=6.0cm]{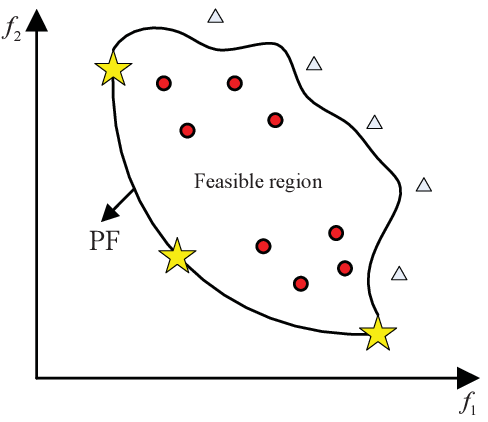}
\par\end{centering}}

\par\end{centering}
\caption{\label{fig:PF1} The PF of  MOP with and without constrains, $m=2$.}
\end{figure}

The PF of an MOP is portrayed both with and without constrains in Fig. \ref{fig:PF1}. Observe from Fig. \ref{fig:PF1} (a) that the Pareto-optimal solutions of the objective functions in the PF (marked as asterisk) provide better values than any other solution in $\mathbf{R}^m$. The ideal solution marked by a square indicates the joint minimum of the objective values $f_1$ and $f_2$ and it is often difficult to reach. The remaining solutions marked as solid circles are all dominated by at least one solution of the PF. In contrast to the unconstrained scenario of Fig. \ref{fig:PF1} (a), in Fig. \ref{fig:PF1} (b), the curve illustrates the PF of a constrained MOP. The solid circles in the feasible region represent the feasible solutions, while the remaining points outside the feasible region (e.g. the points marked by triangles) are infeasible \cite{L2}.

\subsection{MOO Algorithms}

Numerous studies have been devoted to the subject of MOO and a variety of algorithms have been developed for solving MOPs in WSNs. In fact, optimization algorithms are more diverse than the types of objective functions, but the right choice of the objective function has a much more grave impact than the specific choice of the optimization algorithm. Nevertheless, the careful choice of the optimization algorithm is also vital, especially when complex MOPs are considered.

Solving an MOP means finding the PS of the MOP. As mentioned in Section \ref{sec1}, there are various classes of methods designed for generating the PSs of MOPs, such as mathematical programming based scalarization methods, nature-inspired metaheuristics, and so forth. It should be noted that scalarizing an MOP means formulating a single-objective optimization problem whose optimal solutions are also Pareto-optimal solutions to the MOP\cite{Ehrgott2005}. Additionally, it is often required that every Pareto-optimal solution can be reached with the aid of specific parameters of the scalarization.

%%%%%%%%%%%%%%%%%%%%%%%%%%%%%%
%%%%%%%%%%%%%%%%%%%%%%%%%%%%%%

\subsubsection{Mathematical Programming Based Scalarization Methods}
Mathematical programming based classic scalarization methods conceived for MOO include the linear weighted-sum method, the $\epsilon$-constraints method\cite{Ehrgott2005}, and the goal programming (\gls{GP}) based methods\cite{Wierzbicki_1980:Reference_point_method,Marler_Optimization2004,Romero2004:GP,Ozcan2009:GP,
Javier_Sensors2012,Ustun2012:GP_conic_scalarization,Tang2012:GP}, as detailed below.

\paragraph{Linear Weighted-Sum Method}
The linear weighted-sum method scalarizes multiple performance metrics into a single-objective function
by pre-multiplying each performance metric (i.e., component objective) with a weight. Since different performance metrics have different properties and each metric may have a different unit, the normalization must be implemented firstly when using the linear weighted-sum
method for striking compelling performance trade-offs. Then, a different weight is assigned to each metric to get an evaluation function. Finally, the optimal compromise is obtained according to the Pareto-optimal solutions generated by solving multiple single-objective problems, each corresponding to a specific vector of weight values. It can be proved that the optimal solution to each of these single-objective problems is a Pareto-optimal solution to the original multi-objective problem, i.e., the image of these solutions belong to the PF.

The linear weighted-sum method is easy to implement and can avoid complex computations, provided that the weights are appropriately chosen, since only a single optimal value has to be calculated for each single-objective problem. It is worth pointing out that all the weights are in the range $[0, 1]$, and the sum of them is $1$. However, there is no \textit{a priori} correspondence between a weight vector and a solution vector, and the linear weighted-sum method usually uses subjective weights, which often results in poor objectivity and makes the objectives to be optimized sensitive to the weights. The need to solve multiple single-objective optimization problems with the aid of different sets of weight values also implies that a substantial overall computational complexity may be imposed.  Furthermore, the lack of a reasonable weight allocation method degrades its scientific acceptance. To elaborate a little further, typically the decision maker is \textit{a priori} unaware of which weights are the most appropriate ones to generate a satisfactory solution, hence he/she does not know in general how to adjust the weights to consistently change the solution. This also means that it is not easy to develop heuristic algorithms that, starting from certain weights, are capable of
iteratively generating weight vectors to reach a certain portion of the PF.
In addition, the linear weighted-sum method is incapable of reaching the non-convex parts of the PF. Finally, a uniform spread of the weight values, in general, does not produce a uniform spread of points on the PF. This fact implies that usually all the points are grouped in certain parts of the PF,
while some (potentially significant) portions of the PF based trade-off curve have not been
reached.

Note that if the decision maker has \textit{a priori} preference among the multiple objectives considered, or he/she would like to select the most satisfactory solution from the Pareto-optimal solutions obtained, a powerful multiple criteria decision-making method referred to as the analytical hierarchy process (\gls{AHP})\cite{Saaty_2008:AHP_review, Saaty_2008:AHP_book, Ray_2010:AHP,Wang_2014:AHP,Li_2014:AHP} can be used to determine the relative weights. Using AHP, the weights can be flexibly altered according to the specific application requirements. AHP has been widely used in the context of trade-off mechanisms (e.g. \cite{WangICIS2010} and \cite{GaoICSPSS2010}), where AHP first decomposes a complex problem into a hierarchy of simple subproblems, then synthesizes their importance to the original problem, and finally chooses the best solution.

\paragraph{$\boldsymbol {\varepsilon}$-Constraints Method}
The $\boldsymbol {\varepsilon}$-constraints method creates a single-objective function, where only one of the original objective functions is optimized while dealing with the remaining objective functions as constraints. This method can be expressed as \cite{Eichfelder2009}:
\begin{align}
\min &{f_i({\bf x})}  \nonumber\\
\mbox{s.t.}\quad
   &f_j\leq \varepsilon_j,~j\neq i, \nonumber\\
   &H({\bf x})=0, \nonumber\\
   &G({\bf x})\leq 0,
\end{align}
where $f_i({\bf x}),~i=(1,2,\ldots,N)$ is the selected function for optimization and the remaining $N-1$ functions act as constraints. It was proved by Miettinen\cite{Miettinen_1998} that if an objective $f_j$ and a vector $\boldsymbol {\varepsilon} = (\varepsilon_1,...,\varepsilon_{j-1}, \varepsilon_{j+1},...,\varepsilon_N) \in \mathbf {R}^{N-1}$ exist, such that ${\bf x}^*$ is an optimal solution to the above problem, then ${\bf x}^*$ is a weak Pareto optimum of the original MOP. Therefore, this method is capable of obtaining \textit{weak Pareto-optimal} points by varying the $\boldsymbol {\varepsilon}$ vector, but it is not guaranteed to obtain all of them. Under certain stronger conditions, it can even obtain the strict Pareto optimum\cite{Miettinen_1998}. This method is very intuitive and the parameters $\varepsilon_j$ used as upper bounds are easy to interpret. Another advantage of this method is that it is capable of finding Pareto-optimal solutions on a non-convex PF. Similar to the linear weighted-sum method, having to empirically vary the upper bound $\varepsilon_j$ also implies a drawback of the $\boldsymbol {\varepsilon}$-constraints method, and it is not particularly efficient if the number of objective functions is higher than two.

\paragraph{Goal Programming (\gls{GP})}
Instead of maximizing multiple objectives, GP is an analytical approach devised for solving MOPs, where the goal values (or targets) have been assigned to all the objective measures and where the decision-maker is interested in minimizing the ``non-achievement'' of the corresponding goals. In other words, the underlying assumption of GP is that the decision-maker seeks a satisfactory and sufficient solution with the aid of this strategy. GP can be regarded as an extension or generalization of linear programming to handle multiple  conflicting objective measures. Each of these measures is given a goal or target value to be achieved. The sum of undesirable deviations from this set of user-specified target values is then minimized with the aid of a so-called achievement function. There are various forms of achievement functions, which largely determine the specific GP variant. The three oldest and still widely used forms of achievement functions  include the weighted-sum (Archimedean), preemptive (lexicographic) and MINMAX (Chebyshev)\cite{Wierzbicki_1980:Reference_point_method,Marler_Optimization2004,Romero2004:GP,Ozcan2009:GP,
Javier_Sensors2012,Ustun2012:GP_conic_scalarization,Tang2012:GP}.

There exist other scalarization methods devised for MOO, such as the conic scalarization method of\cite{Kasimbeyli2011:conic_scalarization,Ustun2012:GP_conic_scalarization}.
%%%%%%%%%%%%%%%%%%%%%%%%%
%%%%%%%%%%%%%%%%%%%%%%%%%
\subsubsection{Nature-Inspired Metaheuristic Algorithms}
MOPs are more often solved by nature-inspired metaheuristics, such as multi-objective evolutionary algorithms (\glspl{MOEA}) \cite{T9,T3} and swarm intelligence based optimization algorithms (\glspl{SIOA}) \cite{Y2}.  This is because most classical optimization methods are based on a limited number of standard forms, which means that they have to comply with the particular structures of objective functions and constraints. However, in realistic scenarios it is often impossible to accurately characterize the physical problem with an ideal standard-form optimization problem model. Additionally, many complicated factors, such as a large number of integer variables, non-linearities, and so forth may occur. Both of them can make the realistic problems hard to solve. Therefore, the classical mathematical programming based optimization methods may not be suitable for solving the MOPs encountered in real-world WSNs.

Over the most recent decade, metaheuristics have made substantial progress in approximate search methods for solving complex optimization problems \cite{Jones2002}. A metaheuristic technique guides a subordinate heuristic using concepts typically derived from the biological, chemical, physical and even social sciences, as well as from artificial intelligence, to improve the optimization performance. Compared to mathematical programming based methods, metaheuristics based optimization algorithms are relatively insensitive to the specific mathematical form of the optimization problems. However, the higher the degree of accuracy required, the higher the computational cost becomes. So far, the field of metaheuristics based optimization algorithms has been mostly constituted by the family of evolutionary algorithms (\glspl{EA}) \cite{T3} and the family of SIOAs \cite{T41}.
In the following, we will review some of their salient representatives.

\paragraph{Evolutionary Algorithms (EAs)}

EAs belong to the family of stochastic search algorithms inspired by the natural selection and survival of the fittest in the biological world. The goal of EAs is to search for the globally near-optimal solutions by repeatedly evaluating the objective functions or fitness functions using exploration and exploitation methods. Compared with mathematical programming, EAs are eminently suitable for solving MOPs, because they simultaneously deal with a set of solutions and find a number of Pareto-optimal solutions in a single run of the algorithm. Additionally, they are less susceptible to the specific shape or to the continuity of the PF, and they are also capable of approximating  the discontinuous or non-convex PF\cite{T4}. The most efficient PF-based MOEAs have been demonstrated to be powerful and robust in terms of solving MOPs\cite{T3}.

\begin{figure}[t]
\centering
\includegraphics[width=7cm]{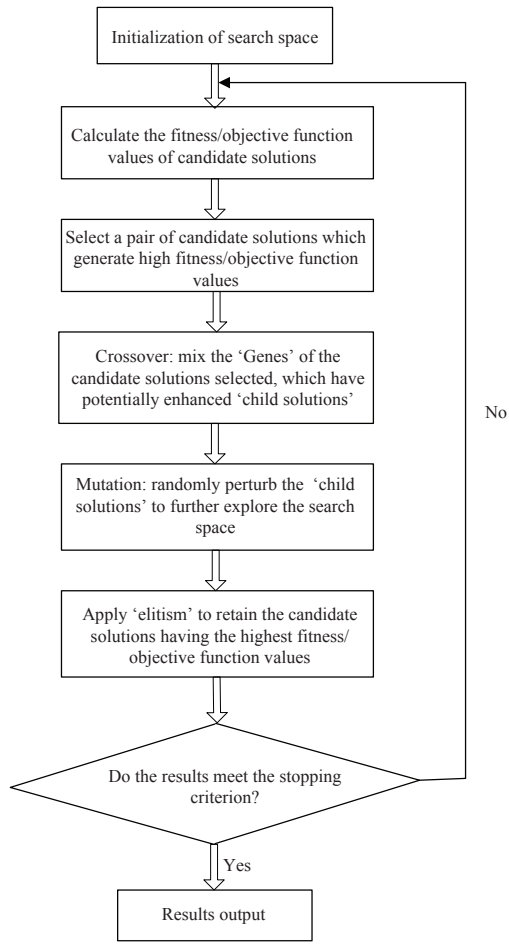}
\caption{Simplified flow-chart of a GA.}
\label{sys10}
\end{figure}

{\textit{$\bullet$ Genetic Algorithms (GAs):}}
GAs constitute the most popular branch of EAs \cite{Q15}.
GAs are based on genetics and evolutionary theory, and they have been successfully used for solving diverse optimization problems, including MOPs. They have appealing advantages over traditional mathematical programming based algorithms \cite{T41} in terms of handling complex problems and convenience for parallel implementation. GA can deal with all sorts of objective functions no matter they are stationary or transient, linear or nonlinear, continuous or discontinuous. These advantageous properties of GAs have inspired their employment in solving MOPs of WSNs. GAs rely on the bio-inspired processes of initialization, evaluation, selection, crossover, mutation, and replacement, as portrayed in its simplest form in the flow-chart of Fig. \ref{sys10}.

The MOGA \cite{Q46} has attracted particularly extensive research attention among all the algorithms of MOO. By operating on the generation-by-generation basis, a number of Pareto-optimal solutions can be found throughout the evolution generations. Thus, obtaining the Pareto-optimal solution set provides us with a set of flexible trade-offs. Several solution methods based on MOGAs were presented in the literature\cite{M1}, \cite{T42} for optimizing the layout of a WSN. More specifically, the authors of \cite{M1} advocated a MOGA for the optimal deployment of static sensor nodes in a region of interest, which simultaneously maximized the coverage area and the network's lifetime. The authors then extended their work to three specific surveillance scenarios in \cite{T42} using the same MOGA. Recently, the non-dominated sorting genetic algorithm (\gls{NSGA})\cite{Q47}, the niched Pareto genetic algorithm (\gls{NPGA}) \cite{T7} and the SPEA \cite{T9} have been recommended as the most efficient MOEAs.

\begin{figure}[t]
\centering
\includegraphics[width=7.5cm]{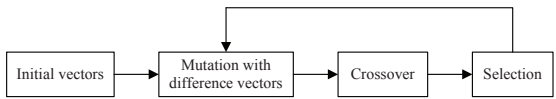}
\caption{Simplified illustration of a DE algorithm.}
\label{DE}
\end{figure}

{$\bullet$ \textit{Differential Evolution (\gls{DE}):}}
DE was developed by Storn and Price \cite{Storn1997}. It is arguably one of the
most powerful real-valued optimization algorithms. DE relies on similar computational steps
as employed by a standard EA.
It commences its operation from randomly initiated parameter vectors, each of which (also called \textit{genome} or \textit{chromosome}) forms a candidate solution to the optimization problem.
Then, a mutant vector (known as \textit{donor vector}) is obtained by the differential mutation operation. To enhance the potential diversity of the candidate solutions, a crossover operation comes into play after generating the \textit{donor vector} through mutation. The final step of the algorithm calls for selection in order to determine whether the target vector survives to the next generation \cite{DasTEC2011}. The main stages of the DE algorithm are illustrated in Fig. \ref{DE}.

However, unlike traditional EAs, the DE algorithm is much simpler to implement,
with only a few parameters to be set. Therefore, DE has drawn much attention and has been successfully applied in numerous domains of science and engineering (see e.g., \cite{Rubio-Largo2013}, \cite{Murugeswari2015}).

%%%%%%%%%%%%%%%%%%%%%%
{$\bullet$ \textit{Artificial Immune System (AIS):}}
AIS is a computational intelligence paradigm inspired by the biological
immune system. It has been applied to a variety of optimization problems and has shown several attractive properties that allow EAs to avoid premature convergence and to enhance local search \cite{Dasgupta2011}.
AIS is capable of recognizing and combating pathogens.
Molecular patterns expressed on those pathogens are referred to as \textit{antigens}. An \textit{antigen} is any
molecule that can be recognized by the immune system and is capable of provoking the immune response. This immune
response is specific to each \textit{antigen}. The cells called \textit{lymphocytes} have a vital role in the immune system. There
are two types of \textit{lymphocytes}: B cells and T cells.
When an \textit{antigen} is detected, B cells that best recognize the \textit{antigen}, will proliferate by
cloning. Some of these new cloned cells will differentiate into plasma cells, which are
the most active \textit{antibody} secretors.
These cloning and mutation processes are termed the clonal selection principle \cite{Coello_EAAI2005},
which is one of the inspiring methodologies employed in AIS for solving optimization problems.
Based on the clonal selection principle, an algorithm
is developed, where various immune system aspects are taken into account, such
as the maintenance of the memory cells, selection and cloning of the most stimulated cells, death of non-stimulated cells, re-selection of the clones with higher affinity, as well as the generation and maintenance of diversity. The steps of the AIS are provided in form of pseudocode in Table \ref{table1}.
\begin{table}
\centering
\extrarowheight 3pt
\caption{Pseudocode of Artificial Immune System.}
\begin{tabular}{|l|}
\hline
 Repeat \\
 ~1. Select an antigen $\mathcal{A}$ from population of antigens; \\
 ~2. Take $R$ antibodies from population of antibodies; \\
 ~3. For each antibody $r\in R$,  \\
 ~~~~match it against the selected antigen $\mathcal{A}$; \\
 ~4. Find the antibody with the highest match score, \\
 ~~~~ break ties randomly, and compute its match score; \\
 ~5. Add match score of winning antibody to its fitness; \\
 Until the maximum number of cycles is reached. \\
\hline
\end{tabular}
\label{table1}
\end{table}

Similar to the computational frameworks of EA, AIS can be readily incorporated into
the evolutionary optimization process and particularly, AIS are often exploited in evolutionary techniques devised for MOO to avoid  premature convergence. On the other hand, the main distinction between the field of AIS and GAs is the nature of population development \cite{Omkar2008}. Specifically, the population of GAs is evolved using crossover and mutation operations. However, in the AIS, similar to evolutionary
strategies where reproduction is a cloning, each child produced by a cell is an exact copy of its parent. Both
algorithms then use mutation to alter the progeny of the cells and introduce further genetic variations.
%%%%%%%%%%%%%%%%%%%%%%%%%%%

\textit{$\bullet$ Imperialist Competitive Algorithm (\gls{ICA}):}
Inspired by the socio-political evolution process of imperialism and imperialistic competition, ICA was originally proposed by Atashpaz-Gargari and Lucas in 2007 \cite{Atashpaz-Gargari2007}. Similar to other EAs, ICA starts with an initial population, with each of them termed a country. The countries can be viewed as population individuals and are basically divided into two groups based on their power, i.e., imperialists (countries with the least cost function value) and colonies.  After forming initial empires, the colonies start moving toward their relevant imperialist. This movement is a simple manifestation of the assimilation policy, which is pursued
by some of the imperialists and results in improvements of the socio-political characteristics, such as
culture, language and economical policy, in the colonies. Then, the imperialistic competition starts among all
the empires. The imperialistic competition will gradually result in an increase of the power
of stronger empires and a decrease in the power of weaker ones. In this process, weak empires will lose their colonies and eventually collapse. In the long run, ICA converges to a state where only a single powerful empire exists in the world and all the other countries are colonies of that empire\footnote{This might not be the case in realistic world, since every empire in the history has a limited life cycle.}. In this state, the best solution of the optimization problem is given when all colonies and the corresponding imperialist have the same cost.

ICA has been successfully applied in numerous single-objective optimization problems \cite{Yousefi2012,Mohammadi-Ivatloo2012}, where most results indicate that it is superior to the GA in terms of both its accuracy and convergence rate.
The basic structure of the multi-objective imperialist competitive algorithm (\gls{MOICA}) is the same as that of the ICA. However,
new methods are developed to determine the imperialist countries, to define the power of the imperialist countries, and to
calculate the total power of empires for imperialistic competition. Selecting imperialists (best countries) from a set of Pareto-optimal
solutions impacts both the coverage and the diversity of solutions.
This impact is more significant when the optimization problem has a high number of objectives.
A novel MOICA was proposed in\cite{Enayatifar2014} for handling node deployment, where both the fast non-dominated sorting and the Sigma method were employed for selecting the best countries as imperialists.

\paragraph{Swarm Intelligence Optimization Algorithms (SIOAs)}

Swarm intelligence constitutes a branch of artificial intelligence (\gls{AI}), and it exploits the collective behavior of self-organized, decentralized systems that rely on a social structure, such as bird flocks, ant colonies and fish schools. These systems consist of low-intelligence interacting agents organized in small societies (also referred to as swarms), exhibiting traits of intelligence, such as the ability of reacting to environmental threats and the decision making capacities. Swarm intelligence has been utilized in the global optimization framework of controlling robotic swarms in \cite{T41}. Three main SIOAs have been developed, namely, ACO \cite{T24}, PSO \cite{Q48} and artificial bee colony (\gls{ABC}) \cite{Karaboga2007}.

\begin{figure*}[t]
\centering
\includegraphics[width=4.5in]{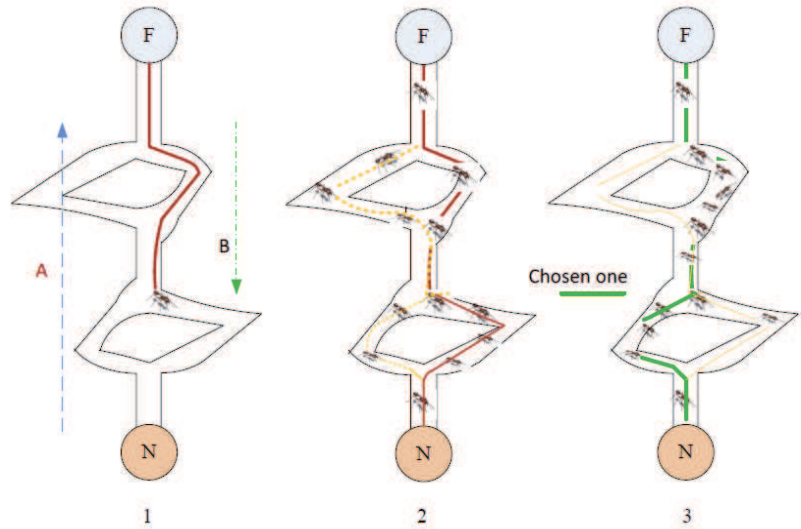}
\caption{A stylized optimization process of ACO.}
\label{sys7}
\end{figure*}

{$\bullet$ \textit{Ant Colony Optimization (ACO):}}
ACO was inspired by the foraging behavior of some ant species. These ants deposit pheromones on the ground in order to mark their nest-to-food paths that should be followed by other members of the colony. Additionally, they also deposit a different type of pheromone to mark dangerous paths for the others to avoid any threat. The ACO algorithm is capable of solving discrete/combinatorial optimization problems in various engineering domains. It was initially proposed by Dorigo in \cite{K1,K2} and has since been widely researched and diversified to solve a class of numerical problems. To illuminate the basic principle of ACO, let us consider the paths $A$ and $B$ of Fig. \ref{sys7} between a nest and a food source as an example \cite{B11}. Furthermore, let us denote by $n_{A(t)}$ and $n_B{(t)}$ the number of ants along the paths $A$ and $B$ at the time step $t$, respectively, and by $P_A$ and $P_B$ the probability of choosing path $A$ and path $B$, respectively. Then, the probability of an ant choosing path $A$ at the time step $t+1$ is given by
\begin{align}
P_A(t+1)=\frac{[c+n_A(t)]^\alpha}{[c+n_A(t)]^\alpha+[c+n_B(t)]^\alpha}
        =1-P_B(t+1),
\end{align}
where $c$ is the degree of attraction of a hitherto unexplored branch and $\alpha$ ($\alpha\geq 0$) is the bias towards using pheromone deposits in the decision process. An ant chooses between the path A or path B using the following decision rule: if $U(0,1) \leq P_A(t + 1)$ then choose path $A$, otherwise choose path $B$. Here, $U$ is a random number having a uniform distribution in the range of $[0, 1]$.
ACO performs well in the dynamic and distributed routing problems of WSNs.
%%%%%%%%%%%%%%%%%%%%%%%%

\begin{figure*}[htb]
\centering
\includegraphics[width=12cm]{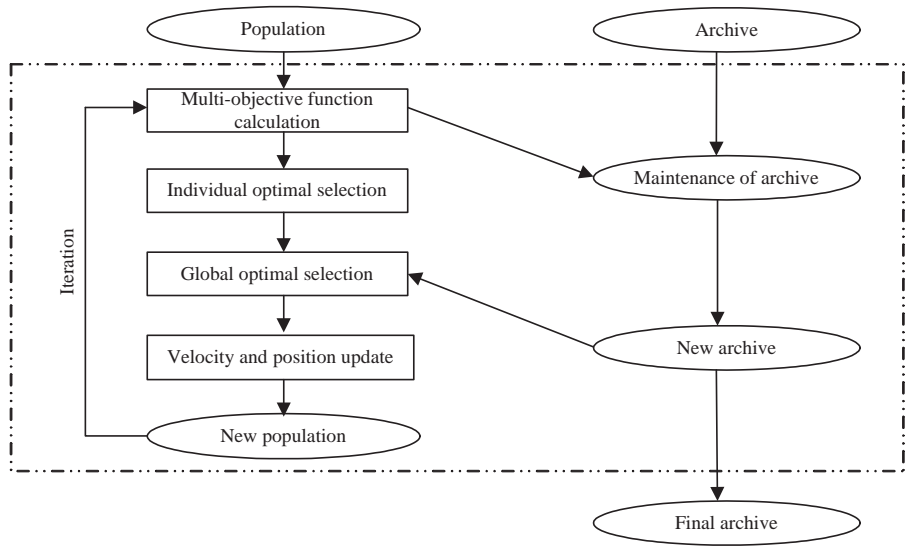}
\caption{The general framework of the multi-objective PSO.}
\label{sys12}
\end{figure*}

{$\bullet$ \textit{Particle Swarm Optimization (PSO):}}
Similar to the underlying philosophy of other swarm intelligence approaches, PSO aims for mimicking the social behavior of a flock of birds. It consists of
a swarm of $s$ candidate solutions, termed particles, which explore an $n$-dimensional hyperspace in search of the global solution. In PSO, the particles regulate their flying directions based both on their own flying experience and on their neighbors' flying experience \cite{K3}. After several improvements conceived by researchers, PSO became an often-used population-based optimizer, which is capable of solving stochastic nonlinear optimization problems at an affordable complexity.
The position of the $i$th particle is represented as $X_{i,d}$, while its velocity is represented as $V_{i,d}$. Each particle is evaluated through an objective function $f(x_1, x_2,...,x_n)$, where we have $f: \mathbb{R}^n \longrightarrow \mathbb{R}$. The cost
(fitness) of a particle close to the global solution is lower (higher) than that of a particle being farther away. The best position of particle $i$ is denoted as
$P_{i,d}$. Then, the particles are manipulated according to the following two equations\cite{K3}:
\begin{align}
V_{i,d}(t+1) & = & wV_{i,d}(t)+\lambda_1r_1(t)[P_{i,d}(t)-X_i(t)] \nonumber\\
             &   & +\lambda_2r_2(t)[P_d(t)-X_i(t)],
\label{equa1}
\end{align}
\begin{align}
X_{i,d}(t+1) & =  X_{i,d}(t)+V_{i,d}(t+1),
\label{equa2}
\end{align}
where we have $1\leq i \leq s$ and $1\leq d \leq n$. Additionally, $\lambda_1$ and $\lambda_2$ are constants, $w$ is the so-called inertia weight, while $P_d$ is the position of the best particle. Still referring to \eqref{equa1}, $r_1(t)$ and $r_2(t)$ are random
numbers uniformly distributed in $[0, 1]$. In the $t$th iteration, the velocity $V$ and the position $X$ are updated using (\ref{equa1}) and (\ref{equa2}). The update process is iteratively repeated until either an acceptable $P_d$ is achieved or a fixed number of iterations $t_{max}$ is reached.
The general framework of the multi-objective PSO is shown in Fig. \ref{sys12}, which includes some key operations, such as the maintenance of the archive, global optimum
selection, as well as the velocity and position update \cite{Sun_Localization2015}. More explicitly, the particle population relies on an archive for storing the Pareto-optimal solutions during the iterative process and for selecting the global
optimum from these solutions. This is the key point in which the multi-objective PSO is different from the traditional single-objective optimization.

The employment of the PSO as a stochastic global optimization algorithm in the MOP of WSNs is relatively new and hence there is a paucity of contributions. A multi-objective routing model based on ACO was proposed in \cite{T24}, which optimizes the network's delay, energy consumption and data packet loss rate. This novel method has been shown to be capable of adapting to different service requirements. The authors of \cite{E4} developed a MOO model based on PSO and fuzzy logic (\gls{FL}) for sensor node deployment, aiming for maximizing the network's coverage, connectivity and lifetime. They have shown that the technique provides efficient and accurate decisions for node deployment in conjunction with low estimation errors.
%%%%%%%%%%%%%%%%%%%%%%%%%%%%

$\bullet$ \textit{Artificial Bee Colony (ABC):}
The ABC algorithm was first introduced by Karaboga and Basturk \cite{Karaboga2007}, and it was derived from the behavior of honey bees in nature. Since the structure of the algorithm is simple, it has been widely used for solving optimization problems. In the ABC model, the position of a food source represents a possible solution to the optimization problem and the amount of nectar in a food source corresponds to the quality (fitness) of the associated solution. The honey bee swarm consists of three groups of bees, namely the employed bees, onlookers and scouts. Correspondingly, the ABC algorithm has three phases\cite{Gutierrez2015,RHLiang2015}. It is assumed that there is only a single artificial employed bee for each food source. Therefore, the number of employed bees in the honey bee swarm is equal to the number of food sources around the hive.

i) Employed bee phase: At the first step, the randomly distributed initial food sources are produced for all employed bees. Then, each employed bee flies to a food source in its memory and determines a neighbor source, whose nectar amount is then evaluated. If the nectar amount of the neighbor source is higher than that of the previous source, the employed bee memorizes the new source position and forgets the old one. Otherwise, it keeps the position of the one in its memory. In other words, an employed bee updates the source position in its memory if it discovers a better food source.

ii) Onlooker phase: After all employed bees have completed the above food-source search process, they return to the hive to share the position and nectar amount of their individual food source with the onlookers. Each onlooker evaluates the nectar information taken from all employed bees and then chooses a food source depending on the nectar amounts of these sources. Therefore, food sources with high nectar content attract a large number of onlooker. Similar to the case of the employed bee, an onlooker then updates the source position in its memory by checking the nectar amount of a neighbor source. If its nectar amount is higher than that of the previous one, the onlooker memorizes the new position and forgets the old one.

iii) Scout phase: As a result, the sources abandoned have been determined, and the employed bee whose food source has been abandoned becomes a scout. New sources are randomly produced by scouts, without considering any experience, in order to replace the abandoned ones.

The above foraging behavior can be simulated using an ABC algorithm to determine the globally optimal solutions of  optimization problems, as shown in\cite{RHLiang2015}.

%%%%%%%%%%%%%%%%%%%%%%%%%%%%%%%%%

\begin{figure}[tbp]
\centering
\includegraphics[width=8.0cm]{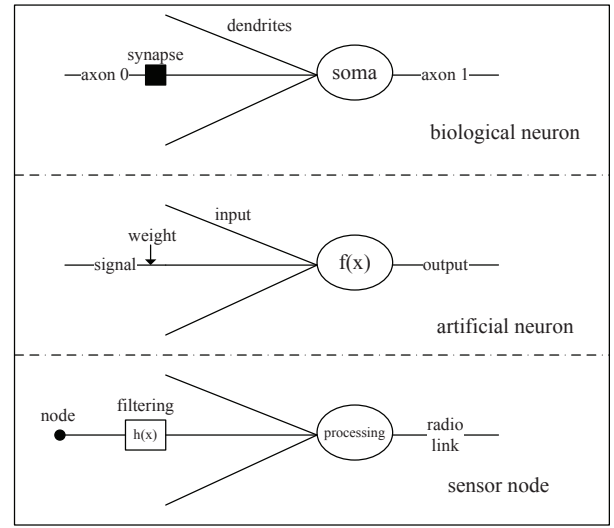}
\caption{The shift procedure from a biological neuron to a sensor node.}
\label{ANN}
\end{figure}

\paragraph{Artificial Neural Network (\gls{ANN})}
ANN is a sophisticated computational intelligence structure inspired by the neurobiological system. It is used for estimating or approximating functions that depend on a large number of inputs that are generally unknown \cite{Barbancho2008}.
The biological neuron consists of dendrites, an axon and a cell body called \textit{soma}. Each neuron may form a connection to another
neuron via the synapse, which is a junction of an axon and a dendrite. The so-called postsynaptic potentials generated within the synapses are received via dendrites and chemically transformed within the \textit{soma}. The axon carries
away the action potential sent out by the \textit{soma} to the next synapse. The analogy of biological neurons to artificial neurons is explained as follows. In artificial neurons, the incoming signals are weighted, which is analogous to what is done in synapses. Then, the weighted signals are further processed. The function $f({\bf x})$ is basically a weighted-sum of all inputs, while the output corresponds to the axon.
In the context of WSNs, the sensor node converts the physical signal to an electronic signal, which is filtered or preprocessed using weighting (analogous to synapse). The subsequent processing within the processor is represented by the particular function $h({\bf x})$, which corresponds to the chemical processing accomplished by the \textit{soma}. Eventually, a sensor node sends out the modified sensor reading via the radio link. This strong analogy shows that the sensor node itself can be viewed as a biological or artificial neuron\cite{Oldewurtel2006}. Therefore, we can readily extend our horizon and regard some WSNs as large-scale ANNs. Having said this, we are fully aware of the inherent dangers of analogies. The shift procedure from a biological neuron to a sensor node is portrayed in Fig. \ref{ANN}.

The entire sensor network can be modelled from an ANN perspective. For each sensor node within the sensor network, we can also rely on ANNs to decide the output action. Thus, it is possible to envision a two-layer ANN architecture for WSNs.
%%%%%%%%%%%%%%%%%%%%%%%%%

\begin{figure}[tbp]
\centering
\includegraphics[width=8.0cm]{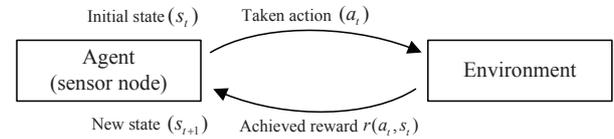}
\caption{Visualization of the $Q$-learning.}
\label{Learning}
\end{figure}

\paragraph{Reinforcement Learning (\gls{RL})}
RL is a powerful mathematical framework that enables an agent (sensor node)
to learn via interacting with its environment and to model a problem as a Markov decision process (\gls{MDP}) \cite{Rovcanin2015}.
The most well-known RL technique is $Q$-learning. The visualization of $Q$-learning is shown
in Fig. \ref{Learning}, where an agent (sensor node) regularly updates its achieved reward based
on the action taken at a given state. The future total reward (i.e., the $Q$-value) of performing an action $a_t$ at state $s_t$ is calculated using
\begin{align}
    Q(s_{t+1},a_{t+1})=Q(s_{t},a_{t})+\lambda \cdot [r(s_{t},a_{t})-Q(s_{t},a_{t})],
\end{align}
where $r(s_{t},a_{t})$ represents the immediate reward of performing
an action $a_t$ at state $s_t$, and $0\leq\lambda\leq1$ is the learning rate that determines how fast learning takes place. This algorithm can be easily implemented in a distributed architecture like WSNs, where each node seeks to choose specific actions that are expected to maximize its long-term rewards. For instance, $Q$-learning has been efficiently used in WSN routing problems \cite{DongGlobecom2007}, \cite{RovcaninAdHoc2014}.
%%%%%%%%%%%%%%%%%%%%%%%%%%%%%%%%
%%%%%%%%%%%%%%%%%%%%%%%%%%%%%%%%

\subsubsection{Other Advanced Optimization Techniques}
There are several other advanced optimization methods capable of achieving appealing performance trade-offs, such as fuzzy logic, game theory, and so forth. Although these methods are less frequently used in WSNs, the trade-offs achieved by them can be compelling.

\paragraph{Fuzzy Logic (FL)}
FL as a mathematical model was introduced by Zadeh in the 1960s \cite{Zadeh1994}.
It is a useful technique that can use human language to describe inputs as well as outputs, and it provides a simple method of achieving a conclusion based on imprecise or ambiguous input information. Since then, the applications of FL have been expanding, especially in adaptive control systems and system identification.

\begin{figure}[tbp]
\centering
\includegraphics[width=8cm]{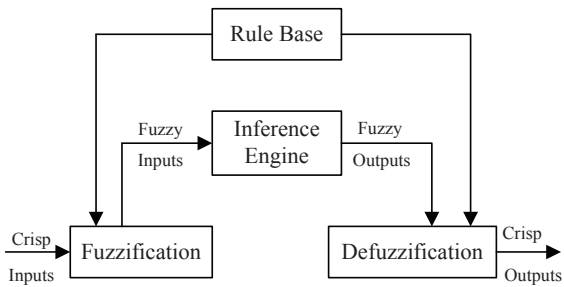}
\caption{Typical structure of the FL.}
\label{fuzzy}
\end{figure}

A fuzzy system comprises four basic elements, namely, fuzzifier, inference engine, fuzzy rule base and defuzzifier\cite{AlShawi2012}, as shown in Fig. \ref{fuzzy}. The fuzzifier converts the inputs into fuzzy variables using membership functions, each of which represents for each object a \textit{degree of belongingness} to a specific fuzzy set. Fuzzy variables provide a mapping of objects to a continuous membership value,  which is normalized in the range $[0, 1]$. Each fuzzy set is represented by a linguistic term, such as ``high'', ``low'', ``medium'', ``small'' and ``large''. The inference engine is often a collection of if-then rules, by which the fuzzy input is mapped to a linguistic output variable according
to the fuzzy rule base. This output variable has to be converted into a crisp output by the defuzzification process, such as the centroid method, averaging method, root sum squared method, and mean of maximum.

Low-complexity FL is suitable for WSNs, and various areas of WSNs have been investigated using the rules of FL. For example, the FL-based routing path search for a maximum network lifetime and minimum delay was investigated in \cite{Gao2016}, where a  fuzzy membership function (edge-weight function) was used for formulating a multi-objective cost aggregation function, which may reflect the effects of all the objectives collectively as a scalar value. As a beneficial result, it offers a beneficial trade-off between maximizing
the network lifetime and minimizing the source-to-sink delay.

\paragraph{Game Theory}
Game theory is a powerful mathematical tool that characterizes the phenomenon of
conflict and cooperation between rational decision-makers \cite{Shisensors2012}. Since game theory introduces a series of successful mechanisms, such as the pricing mechanism, it has achieved a great success in the design of WSNs. In particular,
the pricing schemes can guide the nodes' behaviors towards an efficient Nash equilibrium by introducing a certain degree of coordination into a non-cooperative game.
In \cite{J4}, a Nash equilibrium-based game model, a cooperative coalition game model and an evolutionary game model were used for solving MOPs, respectively.

%%%%%%%%%%%%%%%%%%%%%%%%%%%%

Indeed, a number of MOO approaches have appeared in the literature over the past decade. For the sake of clarity, some representative MOO algorithms are summarized in Table \ref{t1}, including the MOGA\cite{Q46}, NPGA \cite{T7}, NSGA \cite{Q47}, NSGA-II \cite{D13}, SPEA\cite{T9}, the strength Pareto evolutionary algorithm-2 (\gls{SPEA2}) \cite{T10}, the multi-objective messy genetic algorithm (\gls{MOMGA}) \cite{Q41}, the multi-objective messy genetic algorithm-II (\gls{MOMGA-II}) \cite{T12}, the Bayesian optimization algorithm (\gls{BOA}) \cite{T13}, the hierarchical Bayesian optimization algorithm (\gls{HBOA}) \cite{T14}, the Pareto archive evolution strategy (\gls{PAES}) \cite{T15}, the Pareto envelope-based selection algorithm (\gls{PESA}) \cite{T16}, the Pareto envelope-based selection algorithm-II (\gls{PESA-II}) \cite{Q49}, multi-objective differential evolution (\gls{MODE}) \cite{DasTEC2011},  multi-objective evolutionary algorithm based on decomposition (\gls{MOEA/D}) \cite{Q9b}. Additionally, there are some other methods, such as the multi-objective genetic local search (\gls{MOGLS}) \cite{T21}, the multi-objective Tabu search (\gls{MOTS}) \cite{T22}, the multi-objective scatter search (\gls{MOSS}) \cite{T23}, ACO \cite{T24},  PSO \cite{Q48}, ABC \cite{Gutierrez2015}, FL \cite{Gao2016},
%TPSMA \cite{Abidin2014},
ANN, AIS, game theory \cite{J4}, MOICA \cite{Enayatifar2014}, memetic algorithm (\gls{MA}) \cite{Chen_sensors2014}, and centralized immune-Voronoi deployment algorithm (\gls{CIVA}) \cite{Abo-Zahhad2016}, just to name a few.
\begin{table*}\small
\centering
\setlength{\tabcolsep}{2pt}
\renewcommand{\arraystretch}{1.3}
\extrarowheight 3pt
\caption{Qualitative Comparison of Representative MOO Algorithms.}
\begin{tabular}{| p{3cm} | p{2cm} | p{2cm} | p{2cm} | l |}
%\toprule
\hline
\textbf{Approach} & \textbf{Complexity} & \textbf{Convergence} & \textbf{Scalability} & \textbf{Optimality}   \\
\hline
\hline
linear weighted-sum method    & moderate  & fast   & limited   & mathematically guaranteed optimal \\
\hline
$\boldsymbol {\varepsilon}$-constraints method & low       & fast   & limited   & mathematically guaranteed optimal \\
\hline
GP       & moderate  & fast      & good      & mathematically guaranteed optimal \\
\hline
MOGA     & moderate  & fast      & limited   & empirically very near-optimal \\
\hline
NSGA     & high      & slow      & limited   & empirically very near-optimal \\
\hline
NSGA-II  & moderate  & fast      & good      & empirically very near-optimal \\
\hline
NPGA     & low       & slow      & limited   & empirically very near-optimal \\
\hline
SPEA     & high      & fast      & good      & empirically very near-optimal \\
\hline
SPEA2    & high      & fast      & good      & empirically very near-optimal \\
\hline
PAES     & moderate  & fast      & limited   & empirically very near-optimal  \\
\hline
PESA     & moderate  & moderate  & moderate  & empirically very near-optimal \\
\hline
PESA-II  & low       & moderate  & good      & empirically very near-optimal \\

\hline
MOEA/D   & low       & fast      & good      & empirically very near-optimal \\
  \hline
MOGLS    & moderate  & fast      & limited   & empirically very near-optimal \\
 \hline
MOMGA    & high      & moderate  & moderate  & empirically very near-optimal \\
 \hline
MOMGA-II & low       & fast      & good      & empirically very near-optimal \\
  \hline
MOTS     & moderate  & slow      & good      & near-optimal \\
  \hline
MOSS     & moderate  & moderate  & limited   & near-optimal \\
\hline
MODE     & high      & moderate  & limited   & empirically very near-optimal  \\
  \hline
BOA      & high      & slow      & moderate  & near-optimal \\
  \hline
HBOA     & low       & moderate  & limited   & near-optimal \\
\hline
PSO      & low       & slow      & limited   & empirically very near-optimal \\
\hline
ACO      & high      & moderate  & good      & empirically very near-optimal \\
\hline
ABC      & low       & fast      & good      & empirically very near-optimal \\
\hline
FL       & low       & fast      & limited   & empirically very near-optimal   \\
\hline
ANN      & low       & slow      & good      & empirically very near-optimal  \\
\hline
AIS      & moderate  & moderate  & good      & near-optimal \\
\hline
MOICA    & moderate  & fast      & good      & near-optimal \\
\hline
Game Theory & moderate  & low    & good      & empirically very near-optimal \\
\hline
MA       & moderate  & fast      & good      & near-optimal \\
\hline
CIVA     & low       & slow      & good      & near-optimal \\
\hline
RL       & low       & fast      & good      & empirically very near-optimal \\
  \hline
\end{tabular}
\label{t1}
\end{table*}

\subsection{Software Tools}
At the time of writing, numerous software tools are available for solving MOPs. These software packages are briefly introduced in Table \ref{t8}, including BENSOLVE \cite{Lohne2016_European}, the distributed evolutionary algorithms in Python (DEAP) \cite{Fortin2012_JMLR}, Decisionarium \cite{RPH2003_JMCDM}, D-Sight \cite{Hayez2012_IJDSST}, the graphical user interface designed for multi-objective optimization (GUIMOO) \cite{GUIMOO:project_web}, the intelligent decision support system (IDSS) \cite{Lin1996_ICIT}, iSIGHT \cite{Koch2002_Structural}, jMetal \cite{Durillo2010_ICEC}, the multiple objective metaheuristics library in C++ (MOMHLib++) \cite{Ziadloo2009_ICCIMSA}, ParadisEO-MOEO \cite{Liefooghe2007_CS, Liefooghe2011_European}, SOLVEX \cite{Potapov1995_IFIP} and  WWW-NIMBUS \cite{Miettinen2000_Computers}.

\begin{table*}\scriptsize
\centering
\setlength{\tabcolsep}{2pt}
\renewcommand{\arraystretch}{1.3}
\extrarowheight 3pt
\caption{Representative Software Tools.}
\begin{tabular}{| p{1.6cm} | l | p{12cm} |}
%\toprule
\hline
\textbf{Software Tools \,\,(alphabetically)} & \textbf{License} & \textbf{Brief Introduction}   \\
\hline
\hline
BENSOLVE    & open source  & BENSOLVE is a solver for vector linear programs, particularly for the subclass of
multiple objective linear programs, which is based on Benson's algorithm and its extensions. [Online] Available: http://bensolve.org/.\\
\hline
DEAP        & open source  &  The DEAP framework is built with the Python
programming language that provides the essential glue for assembling sophisticated evolutionary computation systems. [Online] Available: http://deap.readthedocs.io/en/master/. \\
\hline
Decisionarium  & open source  & Decisionarium  is the first public site for interactive
multicriteria decision support with tools for individual decision making as well as
for group collaboration and negotiation. Also, Decisionarium offers access to complete e-learning modules based on the use of the software. [Online] Available: www.decisionarium.hut.fi. \\

\hline
D-Sight  & open source & D-Sight developed by Quantin Hayez at the CoDE-SMG laboratory is a relatively new MOO software.
It offers multiple interactive and visual tools that help the decision maker to better understand and manage MOPs.
Compared to the previous software, several functional improvements have been implemented in addition to a modern user interface. [Online] Available: http://aca.d-sight.com/. \\

\hline
GUIMOO    & open source  & GUIMOO is free software for analyzing results in MOPs. It
provides visualization of approximative PFs and metrics for quantitative
and qualitative performance evaluation. [Online] Available: http://guimoo.gforge.inria.fr. \\
\hline
IDSS & open source  &  IDSS is a decision support system that makes the extensive use of AI techniques. The development of the IDSS software package is a primary exploration that puts the decision support method into the context of the real-life world. [Online]  Available: http://idss.cs.put.poznan.pl/site/software.html. \\

\hline
iSIGHT  & commercial   & iSIGHT is a generic software framework for integration, automation, and optimization of design processes, which was developed on the foundation of interdigitation to solve complex problems. [Online] Available: http://www.3ds.com/products-services/simulia/products/isight-simulia-execution-engine/portfolio/. \\

\hline
jMetal  & open source & jMetal is a an object-oriented Java-based framework for solving MOPs using metaheuristics. It is a flexible, extensible, and easy-to-use software package. [Online] Available: http://jmetal.sourceforge.net. \\

 \hline
MOMHLib++    & open source  & MOMHLib++  is a library of C++ classes that implements a number of
multiple objective metaheuristics. Each method only needs the local search operation to be
implemented. [Online] Available: http://home.gna.org/momh/. \\

\hline
ParadisEO-MOEO  & open source   & ParadisEO-MOEO is a white-box object-oriented software framework dedicated to the reusable design of metaheuristics for MOO. Technical details on the implementation of evolutionary MOO algorithms under
ParadisEO-MOEO can be found on the ParadisEO website. [Online] Available: 
http://paradiseo.gforge.inria.fr. \\

\hline
SOLVEX  & open source  & SOLVEX is a FORTRAN library of more than 20 numerical algorithms for solving MOPs. We have both the SOLVEX Windows and the SOLVEX DOC versions. [Online] Available: http://www.ccas.ru/pma/product.htm. \\

\hline
WWW-NIMBUS  & open source & WWW-NIMBUS has been designed to solve differentiable and non-differentiable
MOPs subject to nonlinear and linear constraints with bounds on the variables, and it can also accommodate integer
variables. [Online] Available: http://nimbus.mit.jyu.fi/. \\

\hline
\end{tabular}
\label{t8}
\end{table*}

\section{Existing Literature on Using MOO in WSNs \label{sec6}}
The performance metrics presented in Section \ref{sec2} entail conflicting objectives, e.g., the coverage versus lifetime, and the energy consumption versus delay, etc. Therefore, it is necessary to balance multiple trade-offs efficiently by employing MOO techniques. In this section, we present an overview of the existing contributions that are focused on using MOO in the context of WSNs.
\subsection{Coverage-versus-Lifetime Trade-offs}
The reasons why the coverage and the lifetime of a WSN constitute  conflicting objectives are given as follows. Optimizing the coverage represents the maximization of the proportion of the adequately monitored area relative to the total area. From another perspective, the coverage objective desires having a ``spread-out" network layout, where sensor nodes are as far apart from each other as possible in order to minimize the overlap between their sensing disks. This results in
a large number of relay transmissions taking place at the intermediate sensor nodes, especially for those communicating directly with the base station. Hence, the depletion of energy at these sensor nodes will happen sooner, and the network lifetime will then be shorter. On the other hand, in
order to get a longer lifetime, all the sensor nodes tend to communicate using as few hops as possible (or even communicate directly) with the base station, so that their energy is used for their own data transmission as much as possible. This implies a clustered configuration around the base station, with substantial overlap between sensing disks and yielding a poor coverage performance. Table \ref{t2} shows a summary of the existing major contributions on coverage-versus-lifetime trade-offs.
\begin{table*}\scriptsize
\centering
\setlength{\tabcolsep}{2pt}
\renewcommand{\arraystretch}{1.3}
\extrarowheight 3pt
\caption{Coverage-versus-Lifetime Trade-offs.}
\begin{tabular}{| l | l | p{2.8cm} | l | l | l | p{2cm} | l|}
%\toprule
\hline
 \textbf{Ref.} & \textbf{Technical Tasks} & \textbf{Optimization Objectives} & \textbf{Algorithms} & \textbf{Type of Sensors} & \textbf{Topology} & \textbf{Evaluation Methodology} & \textbf{Scope of Applications} \\
\hline
\hline
\cite{M1}  & deployment & maximize coverage;  maximize lifetime & MOGA & homogeneous-static & flat & experimental trial
&satellite or a high-altitude aircraft \\
\hline
\cite{M2} & deployment & maximize coverage;  maximize lifetime &  MOEA/D & homogeneous-static & flat & simulation &
general-purpose\\
\hline
\cite{Konstantinidis_Computer2011} & deployment & maximize coverage;  maximize lifetime &  MOEA/D & homogeneous-static & flat
& simulation & general-purpose\\
\hline
\cite{E4}  & deployment & maximize coverage;  maximize connectivity;  maximize lifetime & hybrid FL and PSO & heterogeneous-static &
flat& simulation & general-purpose \\

\hline
\cite{D44} & data aggregation & maximize coverage;  maximize lifetime (via minimizing latency) & recursive algorithm & homogeneous-static  & flat & simulation &
densely deployed environment \\

\hline
\cite{Abo-Zahhad2016} & deployment & maximize coverage;  maximize lifetime & CIVA & homogeneous-mobile & flat & simulation & general-purpose\\
 \hline
%\bottomrule
\end{tabular}
\label{t2}
\end{table*}

More specifically, Jourdan $et~al.$ \cite{M1} conceived a MOGA for optimizing the layout of WSNs, i.e., the locations of nodes, by considering both the sensing and communication connectivity requirements. The algorithm aims for maximizing both the coverage and the lifetime of the network, hence yielding a PF from which the network can dynamically choose its most desired solution. Konstantinidis $et~al.$ \cite{M2, Q10} considered optimizing both the locations and the transmit power levels of sensor nodes, i.e., the so-called deployment and power assignment problem (\gls{DPAP}) for maximizing the network coverage and lifetime. Using the MOEA/D of \cite{Q9b}, the multi-objective DPAP was decomposed into a set of scalar subproblems in \cite{M2, Q10}.
By extending \cite{M2}, the authors further addressed the $K$-connected DPAP in WSNs for maximizing the network coverage and lifetime under the $K$-connectivity constraints by using, again, the MOEA/D approach \cite{Konstantinidis_Computer2011}. Furthermore, Rani $et~al.$ \cite{E4} proposed a multi-objective PSO and FL based optimization model for sensor node deployment, which is based on the maximization of the network's coverage, connectivity and lifetime.
Choi $et~al.$ \cite{D44} proposed a randomized $k$-disjoint-sensor selection scheme that traded off the coverage against the data reporting latency, while enhancing the attainable energy efficiency depending on the specific type of applications.
Additionally, a CIVA was proposed for mobile WSNs in \cite{Abo-Zahhad2016} to strike an improved   trade-off between coverage and lifetime.
The CIVA comprises two phases: in the first phase, CIVA controls the locations and the sensing ranges of mobile nodes
to maximize the coverage; in the second phase, CIVA adjusts the transmit power of active/sleep mobile nodes to minimize the number of active nodes.

\subsection{Energy-versus-Latency Trade-offs}
\begin{table*}\scriptsize
\centering
\setlength{\tabcolsep}{2pt}
\renewcommand{\arraystretch}{1.3}
\extrarowheight 3pt
\caption{Energy-versus-Latency Trade-offs.}
\begin{tabular}{| l | p{2cm} | p{3cm} | p{2.7cm} | l | p{1.3cm} | p{1.3cm} | p{3cm}|}
%\toprule
\hline
 \textbf{Ref.} & \textbf{Technical Tasks} & \textbf{Optimization Objectives} & \textbf{Algorithms} & \textbf{Type of Sensors} & \textbf{Topology} & \textbf{Evaluation Methodology} & \textbf{Scope of Applications} \\
\hline
\hline
\cite{C22}  & topology and energy management & minimize energy consumption;  minimize latency; minimize network density & a node wake-up based topology-and-energy-management algorithm and the classic geographical adaptive fidelity algorithm & homogeneous-static & flat  & simulation & general-purpose\\

\hline
\cite{D22}  & localization & minimize energy consumption;  minimize latency & a collision avoidance protocol& homogeneous-static & flat  & simulation & general-purpose\\

\hline
\cite{D11}  & clustering &  minimize energy consumption;  minimize latency & a node wake-up scheme based on an asynchronous wake-up pipeline & homogeneous-static & flat with clustering & simulation
& large-scale WSNs \\

\hline
\cite{F2}  &  data aggregation  & minimize energy consumption; minimize latency & dynamic programming & heterogeneous-static &  flat & simulation &real-time monitoring or mission-critical applications \\

\hline
\cite{C2}  & task allocation & minimize energy consumption; minimize latency & a three-phase heuristic & homogeneous-static & flat & simulation
 &real-time application \\
\hline
\cite{Yao2015TNet}& data aggregation & minimize energy consumption; minimize latency & tabu search and ACO & heterogeneous-static & flat & simulation & large-scale WSNs \\

\hline
\cite {C21}  & data aggregation & minimize energy consumption; minimize latency & re-routing algorithms & homogeneous-static & flat & simulation & large and dense WSNs  \\

\hline
\cite {C24}  & routing & minimize energy consumption; minimize latency & a data dissemination protocol & homogeneous-static & flat & simulation & general-purpose\\
\hline
\cite{D55}  & routing & minimize energy consumption; minimize latency & a cluster-and-chain based energy-delay-efficient routing protocol & homogeneous-static & flat with clustering  & simulation & inhospitable physical environments \\

\hline
\cite{F22}  & deployment & minimize energy consumption; minimize latency & uniform algorithm; cluster algorithm & homogeneous-static
& flat & simulation & harsh environments \\

 \hline
\cite{F21}  & scheduling  & minimize energy consumption; minimize latency & analytical method & homogeneous-static & flat & analytical &  abstract multi-state one- and two-dimensional line WSN \\

\hline
\cite{C5}  & scheduling and MAC & minimize energy consumption; minimize latency & hybrid GA and PSO & homogeneous-static & flat & simulation
& general-purpose\\

 \hline
\cite{C4}  & routing & minimize energy consumption; minimize latency & FL & homogeneous-static & flat & simulation & delay-sensitive WSNs \\

 \hline
\cite{F211}  & clustering & minimize energy consumption; minimize latency & NSGA-II & heterogeneous-static & flat & simulation
  & general-purpose \\

 \hline
\cite{F24}  & data aggregation & minimize energy consumption; minimize latency & energy-efficient minimum-latency data aggregation algorithm & homogeneous-static & flat  & simulation
& general-purpose \\

 \hline
\cite{Ammari2009:non_uniform_energy_depletion, C11}  & data forwarding & minimize energy consumption; uniform battery power depletion; minimize latency & weighted scale-uniform-unit sum algorithm& homogeneous-static & flat  & simulation
& sensing applications \\

 \hline
\cite{Shahraki_Computers2011}  & routing & minimize energy consumption; minimize latency & queue theory & heterogeneous-static & hierarchical  & simulation & general-purpose \\

 \hline
\cite{Yao2015TNet}  & data aggregation & minimize energy consumption; minimize latency; maximize lifetime & centralized and distributed heuristics inspired by techniques developed for a variant of the vehicle routing problem  & heterogeneous-static & flat  & simulation
& general-purpose \\

 \hline
\cite{Suto_Access2015}  & data aggregation and processing & minimize energy consumption; minimize latency & integer programming & homogeneous-static & flat  & simulation
& industrial Internet of Things \\

 \hline
\cite{Dong_Parallel2016}  & data aggregation & minimize energy consumption; minimize latency & queue theory & homogenous-static & flat with clustering  & simulation & sensing applications \\
 \hline
%\bottomrule
\end{tabular}
\label{t3}
\end{table*}

Indeed, minimizing the energy consumption requires transmitting the sensed data over reduced distance in each hop. By contrast, minimizing the delay requires minimizing the number of intermediate forwarders between a source and the sink. This goal may be achieved by maximizing the distance between any pair of consecutive forwarders. Furthermore, a reduced search space for candidate forwarders yields an unbalanced distribution of the data forwarding load among nodes, thus causing a non-uniform depletion of their available energy\cite{Ammari2009:non_uniform_energy_depletion,C11}. Therefore, it is necessary to jointly optimize the network's energy consumption and delay. The energy-versus-latency trade-off related issues have been lavishly documented in various specific WSN scenarios\cite{C22,D22,D11,F2, C2, C21,C24,D55,F22,F21,C5,C4,F211,F24, Ammari2009:non_uniform_energy_depletion, Shahraki_Computers2011, Yao2015TNet, Suto_Access2015, Dong_Parallel2016}.
Table \ref{t3} shows a summary of the existing major contributions to energy-consumption-versus-latency trade-offs.

More specifically, in\cite{C22} the authors studied the energy-latency-density trade-off of WSNs by proposing a topology-and-energy-management scheme, which promptly wakes up nodes from a deep sleep state without the need for an ultra-low-power radio. As a result, the WSN designer can trade the energy efficiency of this sleep state against the latency associated with waking up the node. In addition, the authors integrated their scheme with the classic geographical adaptive fidelity algorithm to exploit excess network density. In \cite{D22} Zorzi $et~al.$ developed the energy-versus-latency trade-offs based on the geographical location of the nodes, and proposed a collision avoidance protocol.
Then, Yang $et~al.$ \cite{D11} designed a node wake-up scheme, namely the so-called ``pipelined tone wake-up'', which struck a balance between the energy savings and the end-to-end delay. This node wake-up scheme was based on an asynchronous wake-up pipeline, where the wake-up procedures overlapped with the packet transmissions. It used wake-up tones that allowed a high duty-cycle ratio without imposing a large wake-up delay at each hop.
Yu $et~al.$ \cite{F2} studied the energy-versus-latency trade-offs using the so-called \textit{data aggregation tree}\footnote{In general, data aggregation tree is interpreted as a tree that aggregates information from multiple sources en route to the sink (or recipient). In a tree-based network, sensor nodes are organized into a tree, where data aggregation is performed at intermediate nodes along the tree and a concise representation of the data is transmitted to the root node. Tree-based data aggregation is suitable for applications that involve in-network data aggregation.}\cite{Q42} in a real-time scenario with a specified latency constraint, and developed algorithms for minimizing the overall energy dissipation of the sensor nodes.
The authors of \cite {C2} presented the first work on energy-balanced task allocation in WSNs where both the time and the energy costs of the computation and communication activities were considered. They explored the energy-versus-latency trade-offs of communication activities over the \textit{data aggregation tree} for modelling the packet flow in multiple-source single-sink communications. A numerical algorithm was conceived for obtaining the exact optimal solution, and a dynamic programming based approximation algorithm was also proposed.

In \cite {C21} Borghini $et~al.$ considered the problem of analyzing the trade-offs between the energy efficiency and the delay for large and dense WSNs. They used an analytical model, which facilitated the comparison of the trade-offs in scenarios employing different deployment-phase protocols, and presented a pair of novel algorithms (i.e., latency-oriented/energy-oriented data aggregation tree construction algorithms), which outperformed the existing ones. In \cite {C24}, Ammari $et~al.$ investigated the energy-versus-delay trade-offs of a WSN by varying the transmission range. Huynh $et~al.$ \cite{D55} proposed a cluster-and-chain based energy-delay-efficient routing protocol for WSNs,  where each $k$-hop cluster uses both cluster-based and chain-based\footnote{Note that sensor nodes are distributed into multiple clusters, and each cluster has a cluster head that aggregates all data sent to it by all its members. Afterwards, cluster heads form multiple binary chains, in which each node communicates with the closest neighbor and takes turns transmitting to the base station.} approaches. Each communication round consisted of a cluster- and chain-formation phase, as well as a data transmission phase.

Furthermore, Moscibroda $et~al.$ \cite{F22} analyzed the energy-efficiency versus propagation-delay trade-offs by defining a formal model, with a particular emphasis on the deployment phase. Specifically, the authors presented two new algorithms, one of which is entirely unstructured, while the other is based on clustering.
Leow $et~al.$ \cite{F21} provided an asymptotic analysis of the transmission delay and energy dissipation of a 2D multi-state WSN, where the sensor nodes were equally spaced in a line or in a square grid. They also discussed the transmission delay-energy trade-offs for the case where the energy transmitted attenuates
according to the inverse second-power pathloss law. As a further development, the authors of \cite{C5} presented a new MOO framework conceived for slot scheduling in many-to-one sensor networks. Two specific optimization objectives were considered in\cite{C5}. The first one was to minimize the energy consumption, while the other was to shorten the total delay.
Minhas $et~al.$ \cite{C4} proposed a routing algorithm based on FL for finding a path that offers a desirable balance between the maximum lifetime (associated with energy consumption) and the minimum source-to-sink delay.

In the same spirit, Cheng $et~al.$ \cite{F211} proposed a MOO framework for cluster-based WSNs. The framework was designed to strike attractive trade-offs between the energy consumption and the duration of the data collection process. The effectiveness of this framework was evaluated with a pair of energy-aware clustering algorithms.
However, clustering techniques typically impose bottlenecks during the data collection process and cause extra delays.
Li $et~al.$ \cite{F24} investigated the trade-offs of data aggregation in WSNs in the presence of interference, and they conceived an energy-efficient minimum-latency data aggregation algorithm, which achieved the asymptotically minimal aggregation latency as well as the desired energy-versus-latency trade-offs.
Ammari \cite{C11,Ammari2009:non_uniform_energy_depletion} proposed a data forwarding protocol for finding the best trade-offs among minimum energy consumption, uniform battery power depletion and minimum delay, which relied on slicing the communication range of the nodes into concentric circular bands. He also conceived a novel approach termed the \textit{weighted scale-uniform-unit sum}, which was used by the source nodes for solving this MOP. Shahraki $et~al.$ \cite{Shahraki_Computers2011} defined a new cost function and developed a new intra-cluster routing scheme for balancing the attainable cluster lifetime against the end-to-end delay between the cluster members and the cluster head. In \cite{Yao2015TNet}, Yao $et~al.$ developed a data collection protocol for balancing the trade-offs between energy-efficiency (associated with lifetime) and delay in heterogeneous WSNs, in which a centralized heuristic was devised for reducing the computational cost and a distributed heuristic was conceived for making the algorithm scalable. Both heuristics were inspired by recent techniques developed
for the so-called ``open vehicle routing problems with time deadlines'', which is mainly studied in operational research.
In \cite{Dong_Parallel2016}, Dong $et~al.$ investigated the trade-offs between energy consumption and transport latency minimization under certain reliability constraints in WSNs. Based on the analysis strategy conceived for satisfying sensing application requirements, they proposed
a data gathering protocol named broadcasting combined with multi-NACK/ACK to strike attractive trade-offs. In \cite{Suto_Access2015}, the authors proposed an energy-efficient and delay-aware wireless computing system for industrial WSNs based smart factories.

\subsection{Lifetime-versus-Application-Performance Trade-offs}

In certain sensor network applications, the specific application's performance strongly depends on the amount of data gathered
from each sensor node in the network. However, higher data rates result in increased sensing and communication costs across the sensor network, as well as in escalating energy consumption and reduced network lifetime\cite{Imon_TNet2015}. Thus, there is an inherent trade-offs between the network lifetime and a specific application's performance, while the latter is often correlated to the rate at which the application can reliably send its data across sensor networks. This problem has been extensively studied in recent years. Table \ref{t4} shows a summary of the existing major contributions to lifetime-versus-application-performance trade-offs.
\begin{table*}\scriptsize
\centering
\setlength{\tabcolsep}{2pt}
\renewcommand{\arraystretch}{1.3}
\extrarowheight 3pt
\caption{Lifetime-versus-Application-Performance Trade-offs.}
\begin{tabular}{| l | l | p{3cm} | p{2.5cm} | l | l | p{1.6cm} | p{3.0cm}|}
%\toprule
\hline
 \textbf{Ref.} & \textbf{Technical Tasks} & \textbf{Optimization Objectives} & \textbf{Algorithms} & \textbf{Type of Sensors} & \textbf{Topology} & \textbf{Evaluation Methodology} & \textbf{Scope of Applications} \\
\hline
\hline
\cite{C6}  &  routing  & maximize lifetime; maximize network utility & subgradient algorithm
 & heterogeneous-static & hierarchical & simulation & self-regulating WSNs \\
\hline
\cite{C7}  & routing & maximize lifetime; maximize network utility & gradient projection algorithm & heterogeneous-static & hierarchical
& simulation & cross-layer applications \\
\hline
\cite{B1}  & routing & maximize lifetime; maximize network utility & subgradient algorithm & homogeneous-static & flat
& simulation & large-scale WSNs \\
\hline
\cite{C9}  & optimal flow control & maximize lifetime; maximize network utility & gradient projection algorithm & heterogeneous-static & flat
& simulation & video technology WSNs \\
\hline
\cite{D0}  & routing & maximize lifetime; maximize aggregate utility & stochastic quasi-gradient algorithm & heterogeneous-static & hierarchical
& simulation & online query applications  \\
\hline
\cite{C10}  & scheduling and MAC & maximize lifetime; maximize throughput & analytic method & homogeneous-static & flat
& simulation & general-purpose \\
\hline
\cite{Xie_SensorsJ2013}  & MAC routing & maximize lifetime; maximize throughput &  improved MAC protocol & homogeneous-static & flat with clustering
& simulation &   information service oriented sensing \\
\hline
\cite{Liao_TVT2015}  & optimal flow control & maximize lifetime; maximize network utility &  distributively extended primal-dual algorithm & homogeneous-static
& flat & simulation & streaming video and audio applications \\
\hline
%\bottomrule
\end{tabular}
\label{t4}
\end{table*}

In \cite{C6}, Nama $et~al.$ investigated the trade-offs between network utility and network lifetime maximization in a WSN. They proposed a general cross-layer optimization-based framework that took into account the associated radio resource allocation issues and designed a distributed algorithm by relying on the so-called dual decomposition\cite{Q12} of the original problem.
Similarly, in \cite{C7}, Zhu $et~al.$ studied the trade-off between network lifetime (associated with energy conservation) and rate-allocation by using the gradient projection\cite{Q13} method. However, no detailed information was provided about how to distributively implement this algorithm in the interest of solving the lifetime-versus-rate-allocation trade-off problem in each layer of the open systems interconnection (\gls{OSI}) model. Zhu $et~al.$ also studied the trade-offs between the network's lifetime and fair rate allocation in the context of multi-path routing sensor networks \cite{B1}, where they formulated an MOP subject to a set of convex constraints. They invoked the NUM framework\cite{Q11} and introduced an adjustable factor to guarantee rate-allocation fairness amongst all sensor nodes.
Chen $et~al.$ \cite{C9} have addressed the utility-versus-lifetime trade-offs with the aid of an optimal flow control in a practical WSN. They formulated the problem as a non-linear MOP subjected to certain constraints and introduced auxiliary variables for decoupling the individual objectives embedded in the scalar-valued multi-objective function. The concept of \textit{inconsistent coordination price}\footnote{Inconsistent coordination
price can be interpreted as the auxiliary variable (Lagrange multiplier) for coordinating the energy consumption among the sensor nodes in the constrained
MOP formulated.} was first introduced for balancing the energy consumption of the sensor node and the gradient projection method\cite{Q13} was adopted for designing a distributed algorithm that is capable of finding the optimal rate allocation.
In \cite{D0}, He $et~al.$ focused on the rate allocation problem in multi-path routing WSNs subjected to time-varying channel conditions with two objectives in mind: maximizing the aggregate utility and prolonging the network's lifetime, respectively. They decomposed the optimization problem with the aid of the classic \textit{Lagrange dual decomposition}\cite{Q12} and adopted the stochastic quasi-gradient algorithm\cite{Q13} for solving the primal-dual problem in a distributed way.
Luo $et~al.$ \cite{C10} have also carried out a systematic study of the trade-offs between the network's throughput and lifetime for WSNs having stationary nodes, where the link transmissions were carefully coordinated to avoid interference. The authors used a realistic interference model based on the SINR for modeling the conflicts to avoid, when scheduling the wireless links' transmissions. Their analytical and numerical results provided novel insights into the interplay among the throughput, lifetime and transmit power.
Xie $et~al.$ \cite{Xie_SensorsJ2013} adopted a specific fairness concept to analyze the performance degradation experienced in multirate WSNs and then took into account the trade-offs between the throughput attained and energy consumption imposed. Eventually, a multirate-supportive MAC protocol was proposed for balancing the throughput versus energy consumption. Liao $et~al.$ \cite{Liao_TVT2015} generalized the NUM model to a multiutility framework using MOO and applied this framework to trade off the network utility against the lifetime in WSNs. An extended Lagrange duality method was proposed, which is capable of converging to a selected Pareto-optimal solution.

\subsection{Trade-offs Related to the Number of Nodes}

Intuitively, deploying more sensor nodes would improve the overall event-detection probability of the system, albeit at the expense of
increasing both the energy consumption and deployment cost. This indicates the trade-offs among multiple conflicting
objectives related to the number of nodes. Table \ref{t5} portrays a number of existing contributions to these trade-offs at a glance.

To elaborate, in \cite{E1}, a pair of multi-objective metaheuristic algorithms (MOEA and NSGA-II) have been used for
solving the WSN's layout problem, determining both the number and the locations of the sensor nodes that formed a WSN, so that
reliable full coverage of a given sensor field was achieved. Specifically, the authors focused their attention on the energy efficiency of the network as well as on the number of nodes, while the coverage obtained by the network was considered as a constraint. In \cite{E3}, Jia $et~al.$ proposed a new coverage control scheme based on an improved NSGA-II using an adjustable sensing radius.
The objective was to find the most appropriate balance among the conflicting factors of the maximum coverage rate, the least
energy consumption, as well as the minimum number of active nodes.
As a further development, Woehrle $et~al.$ \cite{F1} have invoked the MOEA to identify attractive trade-offs between low deployment-cost and highly reliable wireless transmission, i.e., to minimize transmission failure probability at as low deployment-cost as possible.
Cheng $et~al.$ \cite{F212} investigated the trade-offs between the maximum affordable number of nodes and the minimum duration of the data collection process in a delay-aware data collection network by exploiting the concepts of Pareto-optimality.

In \cite{D88}, Rajagopalan employed the evolutionary multi-objective crowding algorithm (\gls{EMOCA}) for solving the sensor placement problem.
There were three objectives: maximizing the probability of global target detection, minimizing the total energy dissipated by the sensor network and minimizing the total number of nodes to be deployed. The MOO approach simultaneously optimized the three objectives and obtained multiple Pareto-optimal solutions.
In \cite{C13}, Aitsaadi $et~al.$ considered a multi-objective combinatorial optimization problem, where a new multi-objective deployment algorithm (\gls{MODA}) was proposed. The optimization objective was to reduce the number of deployed nodes, to satisfy the target quality of monitoring,  to guarantee the network's connectivity and finally to maximize the network's lifetime. In \cite{E2}, Le Berre $et~al.$ formulated an MOP of maximizing three objectives. The first objective was the maximization of the coverage area in real time, the second objective was the maximization of the network's lifetime depending on the coverage, and the final objective was to minimize the number of deployed nodes subject to the connectivity on the network. The solutions found by three different algorithms (i.e. NSGA-II, SPEA2 and ACO) were compared.
Recently, a novel MOICA was proposed for sensor node deployment in \cite{Enayatifar2014}, where the  minimization of the number of active sensor nodes and the maximization of the coverage were jointly considered. The numerical results of \cite{Enayatifar2014} demonstrated that the MOICA was capable of providing more-accurate solutions at a lower computational complexity than the existing methods.

\begin{table*}\scriptsize
\centering
\setlength{\tabcolsep}{2pt}
\renewcommand{\arraystretch}{1.3}
\extrarowheight 3pt
\caption{Trade-offs Related to the Number of Sensor Nodes.}
\begin{tabular}{| l | l | p{3cm} | l | l | l | p{2cm} | l|}
%\toprule
\hline
 \textbf{Ref.} & \textbf{Technical Tasks} & \textbf{Optimization Objectives} & \textbf{Algorithms} & \textbf{Type of Sensors} & \textbf{Topology} & \textbf{Evaluation Methodology} & \textbf{Scope of Applications} \\
\hline
\hline
\cite{E1}  &  deployment & minimize number of nodes; minimize energy consumption & MOEA; NSGA-II &  heterogeneous-static & flat
& experimental trial  & complex and real WSNs \\
\hline
\cite{E3}  & coverage control & minimize number of nodes; minimize energy consumption&  NSGA-II & heterogeneous-static & hierarchical
 & simulation & event detection \\
\hline
\cite{F1}  & deployment & minimize number of nodes; guarantee high transmission reliability & MOEA & homogeneous-static & flat & simulation & general-purpose \\
\hline
\cite{F212} & data aggregation & maximize number of nodes; minimize latency & analytic method & homogeneous-static & flat & simulation & time-sensitive applications \\
\hline
\cite{D88}  &  deployment & minimize number of nodes; minimize energy consumption & EMOCA & homogeneous-static & flat &simulation
& event detection   \\
\hline
\cite{C13}  & deployment & minimize number of nodes; guarantee network connectivity; maximize lifetime & MODA & homogeneous-static & flat &simulation &forest fire detection \\
\hline
\cite{E2}  & deployment & minimize number of nodes; maximize coverage; maximize lifetime & NSGA-II; SPEA2; ACO & homogeneous-static
& flat & experimental trial & general-purpose \\
\hline
\cite{Enayatifar2014} & deployment & minimize number of nodes; maximize coverage & MOICA & homogeneous-static & flat & simulation & densely deployed environment \\
\hline
%\bottomrule
\end{tabular}
\label{t5}
\end{table*}

\subsection{Reliability-Related Trade-offs}

The main objective behind the deployment of WSNs is to capture and transmit pictures, videos and other important data to the sink reliably. These applications require us to maintain a strict QoS guarantee\cite{Senouci_Surveys2015,Dobslaw_II2016}. However, maintaining the QoS during routing hinges on numerous factors, such as the energy status of the nodes in the network, the delay, the bandwidth and the reliability requirements. Hence, sophisticated routing protocols have to take into considerations multiple potentially conflicting factors, which makes the problem even more challenging. Table \ref{t6} shows a summary of the existing contributions to reliability-related trade-offs.

To expound a little further, Miller $et~al.$ \cite{D66} studied the trade-offs amongst the energy, latency and reliability. They conceived a meritorious probability-based broadcast forwarding scheme for minimizing both the energy usage and the latency, whilst improving the reliability. EkbataniFard $et~al.$ \cite{C33} have developed a QoS-based energy-aware routing protocol for a two-tier WSN  from the perspective of MOO. The proposed protocol utilizing the NSGA-II efficiently optimized the QoS parameters formulated in terms of the reliability and end-to-end delay, whilst reducing the average power consumption of the nodes, which substantially extended the lifetime of the network.

A high data rate can be maintained by a link at the expense of a reduced delivery reliability, and/or increased energy consumption, which in turn reduces the network lifetime. Again, there is an inherent trade-off among the data rate, reliability and network lifetime. Although numerous treatises have extensively studied the data rate, reliability and network lifetime in isolation, only a few of them have considered the trade-offs among them. Xu $et~al.$ \cite{D99} jointly considered the rate, reliability and network lifetime in a rigorous framework. They addressed the optimal rate-reliability-lifetime trade-offs under a specific link capacity constraint, reliability constraint and energy constraint. However, the optimization formulation was neither separable nor convex. Hence, a series of transformations have been invoked and then a separable and convex problem was derived. Finally, an efficient distributed subgradient dual decomposition (\gls{SDD}) algorithm was developed for striking an appealing trade-off.
In \cite{X1}, Lu $et~al.$ formulated WSN routing as a fuzzy random multi-objective optimization (\gls{FRMOO}) problem, which
simultaneously considered the multiple objectives of delay, reliability, energy, delay jitter, the interference aspects and the energy balance of a path. They introduced a fuzzy
random variable for characterizing the link delay, link reliability and the nodes' residual energy, with the objective of accurately reflecting the random characteristics in WSN routing. Eventually, a hybrid routing algorithm based on FRMOO was designed.
In \cite{Razzaque_Sensors2011}, Razzaque $et~al.$ proposed a QoS-aware routing protocol for body sensor networks, in which a lexicographic optimization approach was used for trading off the QoS requirements and energy costs.
In \cite{Gutierrez2015}, Lanza-Gutierrez $et~al.$ considered the deployment of energy harvesting relay nodes for resolving the conflict among average energy cost, average sensing area and network reliability. Two multi-objective metaheuristics, i.e., the ABC algorithm and the firefly algorithm (\gls{FA}), were applied for solving the problem, respectively.
Ansari $et~al.$ \cite{Ansari_IET2013} considered the energy consumption, reliability, coverage intensity and end-to-end delay trade-offs based on the location of the nodes, and a new multi-mode switching protocol was adopted.
Liu $et~al.$ \cite{Liu_IJDSN2015} proposed an energy-efficient cooperative spectrum sensing scheme for a cognitive WSN by taking into account the energy consumption and the spectrum sensing performance, both of which were jointly optimized using fast MODE.
Xiao $et~al.$ \cite{Xiao_Parallel2015} proposed a time-sensitive utility model for low-duty-cycle WSNs, where they simultaneously took into account the transmission cost, utility, reliability and latency. Moreover, they designed two optimal time-sensitive utility-based routing algorithms to strike the most appropriate balance among these four metrics.

\begin{table*}\scriptsize
\centering
\setlength{\tabcolsep}{2pt}
\renewcommand{\arraystretch}{1.3}
\extrarowheight 3pt
\caption{Reliability-Related Trade-offs.}
\begin{tabular}{| l | p{1.8cm} | p{3cm} | p{2.7cm} | l | l | p{1.6cm} | p{3.5cm} |}
%\toprule
\hline
 \textbf{Ref.} & \textbf{Technical Tasks} & \textbf{Optimization Objectives} & \textbf{Algorithms} & \textbf{Type of Sensors} & \textbf{Topology} & \textbf{Evaluation Methodology} & \textbf{Scope of Applications} \\
 \hline
 \hline
\cite{D66}  & scheduling and MAC & energy-latency-reliability trade-off & a probability-based broadcast forwarding scheme & homogeneous-static & flat
& simulation & general-purpose  \\
\hline
\cite{C33}  & routing & latency-reliability trade-off & NSGA-II & heterogeneous-static & hierarchical& simulation & real time audio-visual applications \\
 \hline
\cite{D99}  & flow control & rate-reliability-lifetime trade-off & stochastic subgradient algorithm & heterogeneous-static & hierarchical & simulation & WSNs with time-varying channel  \\
  \hline
\cite{X1} & routing & latency-reliability-energy trade-off & hybrid FRMOO and GA & heterogeneous-static & hierarchical& simulation
& agriculture surveillance and building monitoring \\
  \hline
\cite{Razzaque_Sensors2011} & routing & reliability-energy trade-off & lexicographic optimization approach & homogeneous-static & flat & simulation
& human body location \\
  \hline
\cite{Gutierrez2015} & deployment & energy-reliability-sensing area trade-off & ABC and FA  & homogeneous-static & flat& simulation & intensive agriculture \\
  \hline
\cite{Ansari_IET2013} & deployment & energy-reliability-coverage-latency trade-off & multi-mode switching protocol &
homogeneous-static & flat& simulation & general-purpose \\
  \hline
\cite{Liu_IJDSN2015} & spectrum sensing & energy-reliability trade-off & fast MODE &
homogeneous-static & flat& simulation & cognitive WSNs \\
  \hline
\cite{Xiao_Parallel2015} & routing & energy-utility-reliability-latency trade-off & time-sensitive
utility-based routing algorithm & homogeneous-static & flat& simulation & general-purpose \\
  \hline
\end{tabular}
 \label{t6}
\end{table*}

\subsection{Trade-offs Related to Other Metrics}

As mentioned in Section \ref{sec1}, in practice it is unfeasible to jointly satisfy the optimum of several potentially conflicting objectives. To circumvent this dilemma, the concept of \textit{Pareto-optimal} has been widely invoked, resulting in a PF generated by all Pareto-optimal solutions of a MOP, where it is impossible to improve any of the objectives without degrading one or several of the others.  Therefore, according to the needs of decision makers and the actual situation of the WSN considered, efficient routing algorithms are required for finding a satisfactory path in WSNs. Table \ref{t7} summarizes other metrics and their trade-offs.

As seen in Table \ref{t7}, Lozano-Garzon $et~al.$ \cite{D10} proposed a distributed $N$-to-$1$ multi-path routing scheme for a WSN by taking into account the number of hops, the energy consumption and the free space loss\footnote{Note that the concept of free space loss is defined as the ratio of the power radiated by the transmitting antenna over that picked up by the recipient in free space conditions. Free space loss is the basic propagation loss.}, and these three objectives were optimized by the SPEA2\cite{T10} with the aim of using the energy efficiently in the network, whilst reducing the packet-loss rate. Bandyppadhyay $et~al.$ \cite{D77} proposed a transmission scheduling scheme using a collision-free protocol for gathering sensor data. Moreover, they studied diverse trade-offs amongst the energy usage, the sensor density, and the temporal/spatial sampling rates. As a further advance, Rajagopalan $et~al.$ \cite{R4} developed a MOO framework for mobile agent routing in WSNs. The multi-objective evolutionary optimization algorithms EMOCA and NSGA-II were employed to find the mobile agents' routes, aiming for maximizing the total detected signal energy, while minimizing the energy consumption by reducing the hop-length.
In \cite{T24}, Wei $et~al.$ established a multi-objective routing model that relies on the delay, energy consumption, data packet-loss rate as its optimization objectives. By adjusting the specific weight of each function, the algorithm adapts well to various services having different energy cost, delay and packet-loss rate requirements. This protocol was implemented using an advanced ACO algorithm that is based on a cloud model\footnote{In contrast to the ``cloud'' concept related to cloud computing and cloud-based networking,  herein the ``cloud model'' represents an effective tool designed for characterizing the uncertain transformation between a qualitative concept, which is expressed by natural language, and its quantitative expression. It mainly reflects two kinds of uncertainty, such as fuzziness and randomness of the qualitative concept. As a reflection of the randomness and fuzziness, the cloud model constructs a mapping from qualities to quantities.}.

%four

For a typical WSN, the accuracy of an application and the longevity of the network are inversely proportional to each other, which is partially due to the finite energy reserves of the nodes and owing to the desire for applications to have large volumes of fresh data to process. Adlakha $et~al.$ \cite{C14} created a four-dimensional design space based on four independent QoS parameters, namely the accuracy, delay, energy consumption and the node density. In order to achieve an improved accuracy or lifetime, various parameters of the individual techniques can be adjusted. The insights and relationships identified in \cite{C14} were not unique to mobility tracking applications, many potential applications of WSNs requiring a balance amongst the factors of energy consumption, node density, latency and accuracy may also benefit from exploiting the results and trends identified in \cite{C14}. For instance, the energy-density-latency-accuracy (\gls{EDLA}) trade-offs have been studied in the context of WSNs in \cite{C15}. By contrast, Armenia $et~al.$ \cite{F23} introduced a Markov-based modeling of the random routing behavior for evaluating the trade-offs between location privacy and energy efficiency in a WSN. Notably, their approach used the information theoretic concept of \textit{privacy loss.} Both the network security and lifetime have been studied by Liu $et~al.$ in \cite{F30}, where
they proposed a three-phase routing scheme, which is termed security and energy-efficient disjoint routing. Based on the secret-sharing algorithm, this routing scheme dispersively and randomly delivered its source-information to the sink node, ensuring that the network security was maximized without degrading the lifetime of WSNs.
As a further development, Tang $et~al.$ \cite{Tang_Parallel2015} proposed a geography-based cost-aware secure routing protocol to address the conflicting lifetime-versus-security trade-off issue in multi-hop WSNs. The design goal was achieved by controlling energy deployment balance and invoking a random walking routing strategy.
Attea $et~al.$ \cite{Attea2015WPS} studied the MOP of how to optimally divide sensor nodes into multiple disjoint subsets so that two conflicting objectives, namely the network lifetime and coverage probability, can be jointly maximized. Each subset of sensors is required to completely cover a set of targets having known locations. Hence, they formulated a multi-objective disjoint set cover (\gls{DSC}) problem, which was tackled by MOEA/D and NSGA-II.
In \cite{Sengupta_Engineering2013}, Sengupta $et~al.$ employed a novel heuristic algorithm, termed MOEA/D with fuzzy dominance (\gls{MOEA/DFD}), for finding the best trade-offs among coverage, energy consumption, lifetime and the number of nodes, while maintaining the connectivity between each sensor node and the sink node.
In \cite{Wang_ICNC2015}, Wang $et~al.$ quantified the probabilistic performance trade-offs among the network lifetime, end-to-end communication delay and network throughput in real-time WSNs. A heuristic-based multiple-local-search technique was employed for finding the solutions.
Inspired by the concept of \textit{potential field} from
the discipline of physics, Zhang $et~al.$ \cite{Zhang_TMC2016} designed a novel potential-based routing algorithm, known as the integrity and delay differentiated routing for WSNs. The objective was to improve the data fidelity
for high-integrity applications and to reduce the end-to-end delay simultaneously.

\begin{table*}\scriptsize
\centering
\setlength{\tabcolsep}{2pt}
\renewcommand{\arraystretch}{1.3}
\extrarowheight 3pt
\caption{Trade-offs Related to Other Metrics.}
\begin{tabular}{| l | l | p{3cm} | p{2.5cm} | p{2.5cm} | l | p{1.6cm} | l|}
%\toprule
\hline
 \textbf{Ref.} & \textbf{Technical Tasks} & \textbf{Optimization Objectives} & \textbf{Algorithms} & \textbf{Type of Sensors} & \textbf{Topology} & \textbf{Evaluation Methodology} & \textbf{Scope of Applications} \\
\hline
\hline
\cite{D10}  & routing & minimize energy consumption; minimize packet loss; minimize hop count & SPEA2 & homogeneous-static & flat & experimental trial & general-purpose \\
\hline
\cite{D77}  & scheduling & density-energy-throughput-delay-temporal sampling rates-spatial sampling rates trade-off & analytical method &
homogeneous-static & flat with clustering & simulation & general-purpose\\
 \hline
\cite{R4}  & routing & maximize detection accuracy; minimize energy consumption; minimize path loss & EMOCA; NSGA-II & heterogeneous-mobile & hierarchical & simulation & general-purpose  \\
  \hline
\cite{T24}  & routing &  minimize energy consumption; minimize latency; minimize packet loss  & improved ACO & heterogeneous-mobile & flat & simulation & large-scale WSNs  \\
  \hline
\cite{C14}  & deployment & accuracy-delay-energy-density trade-off & analytical method & homogeneous-static & flat & simulation
& general-purpose  \\
  \hline
\cite{C15}  & target tracking & energy-density-latency-accuracy trade-off & n/a & homogeneous; static and mobile
& flat & simulation & adaptive mobility tracking \\
  \hline
\cite{F23} & security & privacy loss and energy efficiency trade-off & analytical method & homogeneous-static & flat & simulation
& data mining systems  \\
  \hline
\cite{F30}  & data aggregation & maximize network security; maximize lifetime & a security and energy-efficient disjoint routing scheme & homogeneous-static & flat & simulation
& densely deployed environment \\
  \hline
\cite{Tang_Parallel2015}  & routing & maximize network security; maximize lifetime & a cost-aware secure
routing protocol & homogeneous-static & flat & simulation
& general-purpose \\
  \hline
\cite{Attea2015WPS} & scheduling & maximize coverage probability; maximize network lifetime & the multi-objective DSC problem formulated was solved using MOEA/D and NSGA-II & homogenous-static & flat & simulation  & large-scale surveillance applications\\
 % \hline
%\cite{Abidin2014} & deployment & maximize coverage and connectivity; minimize energy consumption & TPSMA & homogenous-static & flat & simulation  & target monitoring\\
  \hline
\cite{Sengupta_Engineering2013} & deployment & maximize coverage, minimize energy consumption, maximize lifetime, and minimize the number of nodes & MOEA/DFD & homogenous-static & flat & simulation  & general-purpose\\
  \hline
\cite{Wang_ICNC2015} & deployment & maximize lifetime, maximize throughput, and minimize latency & heuristic-based multiple-local-search& homogenous-static & flat & simulation  & general-purpose\\
  \hline
\cite{Zhang_TMC2016} & routing & maximize data fidelity and minimize latency & an integrity and delay differentiated routing algorithm & homogenous-static & flat & simulation  & integrity-sensitive applications \\
  \hline
\end{tabular}
 \label{t7}
\end{table*}

\section{Open Problems and Discussions \label{sec7}}

Despite the increasing attention paid to the  MOO of WSNs, this research area still has numerous open facets for future work, as discussed below.

Most of the studies investigated the MOO of single-hop transmission, whereas only a limited amount of contributions, such as \cite{J1}, paid attention to multi-hop WSNs. Clearly, multi-hop transmission in energy-limited WSNs is essential for conserving transmission energy and thus for prolonging the network's lifetime. Therefore, it is promising to intensify the research of MOO in the context of multi-hop WSNs.

Sensor nodes may move from one place to another as required by the application or may be displaced by objects (human, animals, $etc$). Hence, the mobility of the nodes has a substantial impact on the network's connectivity. For example, if a data packet is long and the node changes its current location during the packet's forwarding, part of the data may be lost at the receiving node. Similarly, when a node selects a routing path but the nodes in the routing path change their locations, the connectivity between the source nodes and destination nodes will be affected. Therefore, the deployment of nodes in highly
dynamic scenarios requires a deployment approach that equips the network with a self-organizing capability.
Artificial potential field (\gls{APF}) techniques\footnote{The APF techniques mainly rely on force vectors, associated with the obstacles or target positions, which may be
linear or tangential and are generated by a potential functions. The concept of APF can be schematically described as ``the manipulator moves in a field of forces, the
position to be reached is an attractive pole for the end effector and obstacles are repulsive surfaces for the manipulator parts''. It has been widely used for mobile robots \cite{Q14}.} have been applied to the problems of formation control and obstacle avoidance in multi-robot systems \cite{J2}. Since these problems are of similar nature to the deployment problem of sensor nodes, the APF techniques may also be used to devise a deployment approach for WSNs.

Since WSNs are typically deployed in physically open and possibly hostile environments, they will be confronted with security attacks ranging from passive attacks, active attacks, and denial-of-service (\gls{DoS}) attacks. Therefore, their security is one of the imperative aspects in future research. Since the security may gravely affect the network performance, especially during the information exchange phase, designing a secure routing protocol for WSNs is a must. Most of the known routing algorithms assume that the nodes are static and rely on a single path. Once the routing is attacked, the network's performance will be
significantly degraded. Therefore, it is necessary to study multi-route protocols conceived for mobile-node based routing capable of satisfying the security and QoS requirements in any real-time application. The trade-offs between the security and QoS requirements will bring about further new challenges.

Existing contributions often assume that the sensor networks are spread across a two-dimensional plane, but in practice they are indeed of three-dimensional (\gls{3D}) nature. The extension of a 2D network into 3D is both interesting and challenging.
In two-tier WSNs, multiple objectives have to be satisfied by the routing algorithms. The authors of \cite{X1} proposed a
routing solution based on the \textit{fuzzy random expected value} model and the standard deviation model of \cite{Q1}\footnote{In
fact, both random uncertainty and fuzzy uncertainty simultaneously exist in link quality and nodes' residual energy. From the perspective of statistics, fuzzy random expected value reflects the average value of a fuzzy random variable, while the standard deviation reflects the degree measure of deviating from the expected value.} to meet the requirements of different applications of the clustered network advocated. Since the fuzzy random expected value model may become inaccurate in uncertain environments, improved routing model based on MOO is necessitated.  Moreover, due to the limitation of GAs, the distributed solving methods based on local information and on the decomposition theory are expected to be further investigated.

Serious natural disasters, such as sandstorms, tsunamis, landslides etc, have routinely damaged the natural environment and inflicted the loss of human lives. Although WSNs provide a promising solution to realize real-time environment monitoring, numerous issues have to be resolved for their practical implementation. One of the major issues is how to effectively deploy WSNs to guarantee large-area sensing coverage and reliable communication connectivity in hostile propagation scenarios.

In recent years, considering the similarity between multi-objective design and
game theory, the latter has also been employed to solve multi-objective design problems. By analogy, $m$-objective designs can be regarded as $m$-player games. The authors of \cite{J3} introduced game theory and the concept of co-evolution into GAs for the sake of solving the MOPs, which has been shown to perform well. In \cite{J4} a Nash-equilibrium based game model, a cooperative coalition game model and an evolutionary game model were used for solving MOP.
Since EAs are capable of finding the global solution of MOPs with good robustness, while Nash games can be used for conflict resolution and
Stackelberg games for hierarchical design, it is promising to solve MOPs of WSNs by combining a Nash game with EAs or combining a Stackelberg game with EAs.
As an adaptive parameter control method based on sensitivity results, the Pascoletti-Serafini scalarization method\cite{Q2} is a more general formulation relying on an unrestricted search direction and an auxiliary vector variable. It has been used for both linear and nonlinear MOPs\cite{Q2}. Naturally, this method can also be applied to solve MOPs in WSNs, yielding approximate solutions of the problems considered. Indeed, for the MOP of multicell networks \cite{J5}, the relationship between the parameters and the optimal solutions was elucidated by the Pascoletti-Serafini scalarization.

The mutually interfering networks are ubiquitous, hence finding innovative cross-layer and cross-network solutions becomes essential. To this end, we believe that many hybrid computational intelligence algorithms which combine the benefits of two or more algorithms should be given careful attention, such as swarm-FL control, neuro-FL control, GA-PSO, GA-ANN, and neuro-immune systems, etc.

Cognitive radio (\gls{CR}) is an emerging wireless communication paradigm, in which the transceivers are capable of intelligently detecting in their vicinity which specific communication channels are in use and which are not. Then, they promptly switch to vacant channels while avoiding occupied ones. This is essentially a form of dynamic spectrum access (\gls{DSA}) \cite{Hattab2014Proceedings}, which may substantially improve the exploitation of the available wireless spectrum. Typically, a transceiver in CR may be capable of determining its geographic location, identifying and authorizing its users, sensing neighboring wireless devices, and automatically adjusting its transmission and reception parameters to allow more concurrent wireless communications in a given spectrum band at a specific location. Depending on which parts of spectrum are available for the operation of CR, we have CR operating either in licensed bands, or in unlicensed bands.  On the other hand, WSNs often use the unlicensed ISM band for communications, but with the rapidly increasing demand of the Internet of Things (\gls{IoT}) based applications (e.g., healthcare and tele-medicine), the currently available ISM band may become insufficient, which can result in various technical problems, such as unreliable transmission of useful data. Therefore, in order to alleviate this ``spectrum crunch'', an emerging trend in WSNs is to equip the wireless sensor nodes with the CR based DSA capability, thus
giving birth to CR aided WSNs (\glspl{CR-WSN}) \cite{Akan2009Network,Ahmada2015Survey,Bukhari2016Survey}.
Due to its potential advantages, CR-WSN might be a promising solution for some specific WSN applications, such as indoor sensing, multiclass heterogeneous sensing, and real-time surveillance \cite{Akan2009Network}. Additionally, WBAN, which is a promising technology for ubiquitous health monitoring systems, is also an application area of CR-WSN. In general, CR-WSNs constitute an  unexplored field with only a handful of studies. More specifically, the authors of \cite{Oto2012TWC} determined the optimal packet size that maximizes the energy-efficiency of a practical realization of a CR-WSN. In
\cite{Ozger2015Mobile}, the authors proposed a spectrum-aware clustering protocol to address the event-to-sink communication coordination issue in mobile CR-WSNs. A cross-layer framework that employed CR to circumvent the hostile propagation conditions for the smart grid was discussed in \cite{Shah2013Industrial}, and the MAC-layer delay of a cognitive sensor node was modeled in \cite{Esmaeelzadeh2016Hoc}. The realization of CR-WSN primarily requires an efficient spectrum management framework for regulating the DSA of densely deployed resource-constrained sensor nodes. Therefore, MOO techniques invoked for designing CR-WSNs should be sufficiently intelligent to differentiate between the traffic types and to satisfy their QoS requirements. The current research efforts on MOPs for large-scale CR-WSNs have to be strengthened.

WSNs play a key role in creating a highly reliable and self-healing smart electric power grid that rapidly responds to online events with appropriate actions. However, due to the  broadcast nature of radio propagation and varying spectral characteristics, establishing a secure and robust low-power smart grid over the WSN must be addressed.
Other technical challenges of WSNs in the smart grid include harsh environmental conditions, tight reliability and latency requirements,
as well as low packet errors and variable link capacity. Until now, there have been only a few approaches available, and more studies are needed in these areas.

\section{Conclusions \label{sec8}}

In this paper, we have provided a tutorial and survey of the research of MOO in the context of WSNs. We commence with the rudimentary concepts of WSNs and the optimization objectives in WSNs, then focus on illuminating the family of algorithms for solving MOPs. Since having multiple objectives in a problem gives rise to a set of Pareto-optimal solutions instead of a single globally optimal solution, none of these Pareto-optimal solutions can be considered to be better than the others on the Pareto front without any further information. Thus the MOO algorithms may be invoked for finding as many Pareto-optimal solutions as possible. Additionally, diverse design trade-offs relying both on classical optimization methods and on the recent advances of MOO have been reviewed in the context of WSNs. Future
research directions on MOO conceived for WSNs include multi-hop transmissions, the deployment of nodes in highly dynamic scenarios, secure multi-path routing protocols and solving optimization problems in 3D networks, CR-WSNs and smart grid.

%\bibliographystyle{IEEEtran}
%\bibliography{./bib/bib}
\bibliographystyle{IEEEtran}
\bibliography{refs}

\begin{IEEEbiography}[{\includegraphics[width=1in,height=1.25in,clip,keepaspectratio]{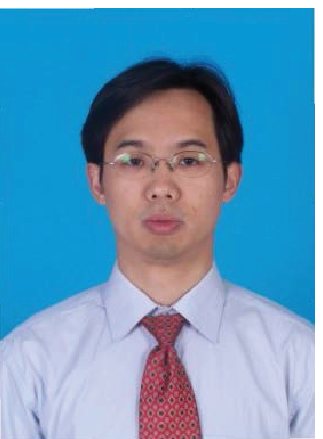}}]{Zesong Fei}(M'07-SM'16) 
received the Ph.D. degree in Electronic Engineering in 2004 from Beijing Institute of Technology (BIT). He is now a Professor with the Research Institute of Communication Technology of BIT, where he is involved in the design of the next generation high-speed wireless communication systems. His research interests include wireless communications and multimedia signal processing. He is a principal investigator of projects funded by National Natural Science Foundation of China. He is also a senior member of Chinese Institute of Electronics and China Institute of Communications.
\end{IEEEbiography}

\begin{IEEEbiography}[{\includegraphics[width=1in,height=1.25in,clip,keepaspectratio]{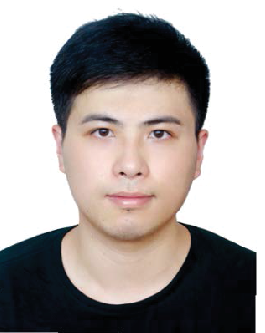}}]{Bin Li}
received the M.S. degree in Communication and Information Systems from Guilin University of Electronic Technology, Guilin, China, in 2013. From Oct. 2013 to Jun. 2014, he was a research assistant in the Department of Electronic and Information Engineering, Hong Kong Polytechnic University, Hong Kong. He is currently pursuing the Ph.D. degree in the School of Information and Electronics, Beijing Institute of Technology, Beijing, China. His research interests include wireless sensor networks, cognitive radio, wireless cooperative networks, physical layer security and MIMO techniques.
\end{IEEEbiography}

\begin{IEEEbiography}[{\includegraphics[width=1in,height=1.25in,clip,keepaspectratio]{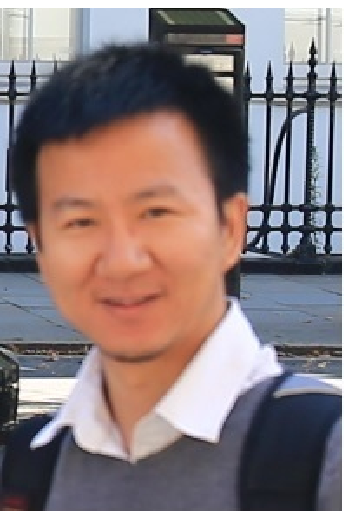}}]{Shaoshi Yang}
(S'09-M'13) received his B.Eng. degree in Information Engineering from Beijing University of Posts and Telecommunications (BUPT), China in 2006, his first Ph.D. degree in Electronics and Electrical Engineering from University of Southampton, U.K. 
in 2013, and his second Ph.D. degree in Signal and Information Processing from BUPT in 2014. He is now a Research Fellow in University of Southampton. From 2008 to 2009, he was an Intern Research 
Fellow with Intel Labs China, working on the mobile WiMAX standardization. His research interests include high-dimensional signal processing for communications, green radio, heterogeneous networks, cross-layer interference management, mathematical optimization and its applications. He received the prestigious Dean's Award for Early Career Research Excellence at University of Southampton, the PMC-Sierra Telecommunications Technology Paper Award at BUPT, and the Best PhD Thesis Award of BUPT. He is a junior member of Isaac Newton Institute for Mathematical Sciences, Cambridge University, and a Guest Associate Editor of IEEE Journal on Selected Areas in Communications. ({http://sites.google.com/site/shaoshiyang/}) 
\end{IEEEbiography}

\begin{IEEEbiography}[{\includegraphics[width=1in,height=1.25in,clip,keepaspectratio]{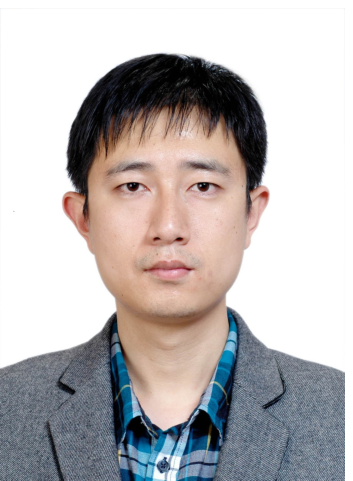}}]{Chengwen Xing}(S'08-M'10) 
received the B.Eng. degree from Xidian University, Xi'an, China, in 2005 and the Ph.D. degree from University of Hong Kong, Hong Kong, in 2010. Since Sept. 2010, he has been with the School of Information and Electronics, Beijing Institute of Technology, Beijing, China, where he is currently an Associate Professor. From Sept. 2012 to Dec. 2012, he was a visiting scholar at University of Macau. His current research interests include statistical signal processing, convex optimization, multivariate statistics, combinatorial optimization, massive MIMO systems and high frequency-band communication systems. Dr. Xing is currently serving as an Associate Editor for IEEE Transactions on Vehicular Technology, KSII Transactions on Internet and Information Systems, Transactions on Emerging Telecommunications Technologies, and China Communications. 
\end{IEEEbiography}

\begin{IEEEbiography}[{\includegraphics[width=1in,height=1.25in,clip,keepaspectratio]{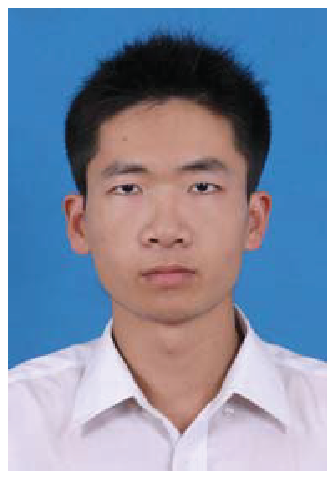}}]{Hongbin Chen}
was born in Hunan Province, China in 1981. He received the B.Eng. degree in Electronic and Information Engineering from Nanjing University of Posts and Telecommunications, China, in 2004, and the Ph.D. degree in Circuits and Systems from South China University of Technology, China, in 2009. He is currently a Professor in the School of Information and Communication, Guilin University of Electronic Technology, China. From Oct. 2006 to May 2008, he was a Research Assistant in the Department of Electronic and Information Engineering, Hong Kong Polytechnic University, China. From Mar. 2014 to Apr. 2014, he was a Research Associate in the same department. His research interests lie in energy-efficient wireless communications.
\end{IEEEbiography}

\begin{IEEEbiography}[{\includegraphics[width=1in,height=1.25in,clip,keepaspectratio]{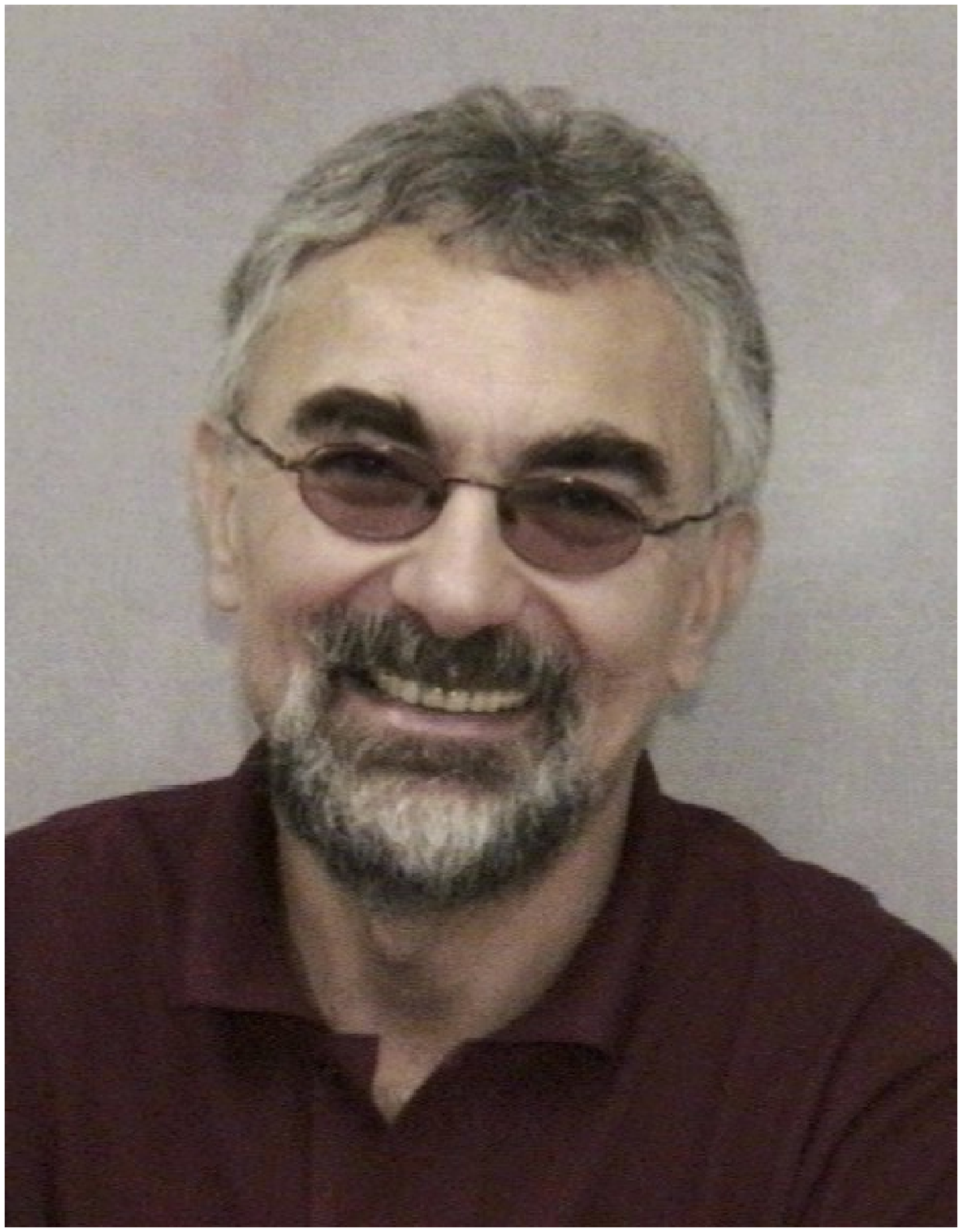}}]{Lajos Hanzo} (M'91-SM'92-F'04) 
(F-REng, F-IEEE, F-IET, F-EURASIP, DSc) received his degree in electronics in
1976 and his doctorate in 1983.  In 2009 he was awarded an honorary
doctorate by the Technical University of Budapest and in 2015 by the
University of Edinburgh.  In 2016 he was admitted to the Hungarian
Academy of Science. During his 40-year career in telecommunications he
has held various research and academic posts in Hungary, Germany and
UK. Since 1986 he has been with the School of Electronics and
Computer Science, University of Southampton, UK, where he holds the
chair in telecommunications.  He has successfully supervised 111
PhD students, co-authored 20 John Wiley/IEEE Press books on mobile
radio communications totalling in excess of 10 000 pages, published
1100+ research contributions at IEEE Xplore, acted both as TPC and General
Chair of IEEE conferences, presented keynote lectures and has been
awarded a number of distinctions. Currently he is directing a
60-strong academic research team, working on a range of research
projects in the field of wireless multimedia communications sponsored
by industry, the Engineering and Physical Sciences Research Council
(EPSRC) UK, the European Research Council's Advanced Fellow Grant and
the Royal Society's Wolfson Research Merit Award.  He is an
enthusiastic supporter of industrial and academic liaison and he
offers a range of industrial courses.  He is also a Governor of the
IEEE VTS.  During 2008 - 2012 he was the Editor-in-Chief of the IEEE
Press and also a Chaired Professor at Tsinghua University, Beijing. Lajos has 25 000+ citations.
For further information on research in progress and associated
publications please refer to http://www-mobile.ecs.soton.ac.uk.
\end{IEEEbiography}

\end{document}